\documentclass[english]{IEEEtran}
\usepackage[T1]{fontenc}
\usepackage[latin9]{inputenc}
\usepackage{xcolor}
\usepackage{pdfcolmk}
\usepackage{float}
\usepackage{amsmath}
\usepackage{amsthm}
\usepackage{amssymb}
\usepackage{graphicx}
\PassOptionsToPackage{normalem}{ulem}
\usepackage{ulem}

\makeatletter

\providecommand{\tabularnewline}{\\}
\floatstyle{ruled}
\newfloat{algorithm}{tbp}{loa}
\providecommand{\algorithmname}{Algorithm}
\floatname{algorithm}{\protect\algorithmname}
\providecolor{lyxadded}{rgb}{0,0,1}
\providecolor{lyxdeleted}{rgb}{1,0,0}

  \theoremstyle{definition}
  \newtheorem{example}{\protect\examplename}
  \theoremstyle{remark}
  \newtheorem{rem}{\protect\remarkname}
  \theoremstyle{plain}
  \newtheorem{lem}{\protect\lemmaname}
  \theoremstyle{plain}
  \newtheorem{thm}{\protect\theoremname}

\usepackage{cite}

\newtheorem{assumption}{Assumption}

\usepackage{subfig}

\makeatother

\usepackage{babel}
\providecommand{\examplename}{Example}
\providecommand{\lemmaname}{Lemma}
\providecommand{\remarkname}{Remark}
\providecommand{\theoremname}{Theorem}

\begin{document}

\title{Stochastic Successive Convex Optimization for Two-timescale Hybrid
Precoding in Massive MIMO}

\author{{\normalsize{}An Liu$^{1}$, }\textit{\normalsize{}Senior Member,
IEEE}{\normalsize{}, Vincent Lau$^{1}$,}\textit{\normalsize{} Fellow,
IEEE}{\normalsize{} and Min-Jian Zhao, }\textit{\normalsize{}Member,
IEEE}{\normalsize{}$^{2}$\\$^{1}$Department of Electronic and Computer
Engineering, Hong Kong University of Science and Technology\\$^{2}$College
of Information Science and Electronic Engineering, Zhejiang University\vspace{-0.3in}
}}
\maketitle
\begin{abstract}
Hybrid precoding, which consists of an RF precoder and a baseband
precoder, is a popular precoding architecture for massive MIMO due
to its low hardware cost and power consumption. In conventional hybrid
precoding, both RF and baseband precoders are adaptive to the real-time
channel state information (CSI). As a result, an individual RF precoder
is required for each subcarrier in wideband systems, leading to high
implementation cost. To overcome this issue, two-timescale hybrid
precoding (THP), which adapts the RF precoder to the channel statistics,
has been proposed. Since the channel statistics are approximately
the same over different subcarriers, only a single RF precoder is
required in THP. Despite the advantages of THP, there lacks a unified
and efficient algorithm for its optimization due to the non-convex
and stochastic nature of the problem. Based on stochastic successive
convex approximation (SSCA), we propose an online algorithmic framework
called SSCA-THP for general THP optimization problems, in which the
hybrid precoder is updated by solving a quadratic surrogate optimization
problem whenever a new channel sample is obtained. Then we prove the
convergence of SSCA-THP to stationary points. Finally, we apply SSCA-THP
to solve three important THP optimization problems and verify its
advantages over existing solutions.
\end{abstract}


\section{Introduction}

Massive MIMO can significantly improve the spectrum efficiency of
wireless systems. The conventional fully digital precoding requires
one RF chain for each antenna, and thus induces huge hardware cost
and power consumption for massive MIMO. As a result, hybrid precoding,
where a high-dimensional RF precoder is connected to a low-dimensional
baseband precoder with a limited number of RF chains, has been proposed
to reduce the hardware cost and power consumption of massive MIMO
base station (BS). 

The early works on this topic focus on studying fast-timescale hybrid
precoding (FHP), where both RF and baseband precoders are adaptive
to the real-time channel state information (CSI). For example, in
\cite{Ayach_TWC14_mmRFprecoding}, a sparse precoding and combining
algorithm based on orthogonal matching pursuit is proposed for single-user
mmWave systems. In \cite{Liang_WCL2014_HBF}, a low-complexity FHP
scheme for multiuser massive MIMO systems is proposed. A limited feedback
hybrid precoding scheme is also proposed in \cite{Heath_TWC2015_HBF}
for multi-user mmWave systems. One disadvantage of FHP is that the
number of RF precoders has to increase with the number of subcarriers
in wideband systems such as orthogonal frequency-division multiple
access (OFDMA) system (since the CSI is different on different subcarriers),
leading to high implementation cost \cite{Liu_TSP2016_CSImassive,Park_TSP17_THP}.
Moreover, as the real-time full CSI is required at the BS, the CSI
signaling overhead is large. 

To overcome the above disadvantages of FHP, a two-timescale hybrid
precoding (THP) scheme is proposed in \cite{Liu_TSP14_RFprecoding,Liu_TSP15_twostageCE}.
In THP, the RF precoder is adaptive to the channel statistics\footnote{In this paper, channel statistics refers to the moments or distribution
of the channel fading realizations.} to achieve the array gain, and the baseband precoder is adaptive
to the low-dimensional effective channel to achieve the spatial multiplexing
gain. THP has several advantages. Since the channel statistics are
approximately the same on different subcarriers \cite{Sadek_TOC08_chstatistic},
THP only needs one RF precoder to cover all subcarriers, which significantly
reduces the hardware cost. Moreover, THP reduces the CSI signaling
overhead because it does not require knowledge of the real-time high-dimensional
CSI. Therefore, THP can achieve a better tradeoff between the implementation
cost and performance, making it more attractive in practice \cite{Liu_TSP2016_CSImassive,Park_TSP17_THP}.

The optimal THP design depends on the RF precoding structure and the
specific application scenario. There are two major RF precoding structures:
the \textit{fully-connected structure} where each antenna is connected
to all the RF chains, and the \textit{partially-connected structure}
where each antenna is only connected to a single RF chain \cite{Molisch_Csurvey2016_HBF}.
For each structure, there are two methods to implement the dynamic
RF precoder. In the codebook-based method, the RF precoder is chosen
from a pre-determined codebook \cite{Ayach_TWC14_mmRFprecoding,Heath_TWC2015_HBF,Liu_TSP14_RFprecoding},
while in the dynamic-phase-shifter-based (DPS-based) method, the phase
of each element of the RF precoder can be quantized and adjusted individually
\cite{Liang_WCL2014_HBF,Park_TSP17_THP}. Under different RF precoding
structures/implementations, the constraint on the RF precoder is different
and thus the optimal THP design is also different. Moreover, in different
application scenarios, the optimization objectives can also be quite
different. For example, for best-effort services, we may want to maximize
the throughput or proportional fairness (PFS) utility under a total
power constraint. For applications with a fixed throughput requirement,
such as video streaming, we may want to minimize the transmit power
subject to an individual throughput requirement. Therefore, it is
important to develop a systematic solution framework to optimize the
THP design for a wide range of applications (i.e., with a general
objective function) under different RF precoding structure/implementation
constraints. 

Unfortunately, the optimization of THP is quite challenging due to
the non-convex stochastic optimization problem involved. The existing
solutions are usually heuristic and only suitable for one application
scenario under a specific RF precoding structure constraint. For example,
the minimum weighted throughput maximization (MWTM) problem is approximately
solved in \cite{Liu_TSP14_RFprecoding} under the fully-connected
and DFT-based RF precoding structure (i.e., the RF precoding codebook
forms a DFT matrix). Specifically, the average data rate is first
replaced with a closed-form lower bound based on knowledge of the
channel covariance matrices, and then the resulting approximate problem
is solved by semidefinite relaxation (SDR). However, the lower bound
may become loose when different user clusters have overlapped angle
of departure (AoD) intervals or the SNR is low. In \cite{Liu_TSP2016_CSImassive},
the average sum-rate maximization problem is solved using the sample
average approximation (SAA) method, again under the fully-connected
and codebook-based RF precoding structure. However, it is known that
SAA has high complexity and is not suitable for online implementation.
The Signal-to-leakage-and-noise ratio (SLNR) maximization problem
is considered in \cite{Park_TSP17_THP} under the fully-connected
and DPS-based RF precoding structure. But the algorithm in \cite{Park_TSP17_THP}
does not consider the fairness issue. 

Note that all the above methods are \textit{offline methods}, which
require a \textit{channel sample collection phase} to collect a large
number of channel samples (to estimate the channel covariance matrices
or construct the SAA functions) before computing the optimized hybrid
precoder, and thus its performance at the channel sample collection
phase is limited. In \cite{Yang_TSP2016_SSCA}, a best-response-based
(BRB) algorithm is proposed for solving general stochastic non-convex
multi-agent optimization problems. The BRB algorithm is an online
algorithm based on stochastic successive convex optimization (SSCA).
However, it only works when the objective function contains expectations
but the constraint can be represented by a deterministic convex set.
In many application scenarios, such as the MWTM problem considered
in \cite{Liu_TSP14_RFprecoding}, there are \textit{stochastic non-convex
constraints} (i.e., the constraint functions are also non-convex and
involve expectations over the random states) involved, which are difficult
to deal with. 

In this paper, we propose an online algorithmic framework called SSCA-THP
to solve a general THP optimization problem without explicit knowledge
of channel statistics. The main contributions are summarized as follows.
\begin{itemize}
\item \textbf{A general THP optimization formulation:} We propose a general
THP optimization formulation with a general smooth objective function,
which can be applied to various application scenarios under different
RF precoding structure/implementation constraints. 
\item \textbf{An online algorithmic framework based on SSCA and its convergence
proof:} We propose an online algorithmic framework called SSCA-THP
to solve the general THP optimization problem with stochastic non-convex
constraints, and establish its convergence to stationary points. At
each iteration of SSCA-THP, quadratic surrogate functions are constructed
for both objective and constraint functions based on a new channel
realization and the current iterate. Then the next iterate is updated
by solving the resulting quadratic optimization problem using a low-complexity
Lagrange dual method. SSCA-THP has several advantages over existing
algorithms. First, it is an online algorithm, meaning that the RF
precoder is updated whenever a (potentially outdated) channel sample
is obtained. As a result, it can achieve a better overall performance
than the offline counterpart. Second, it only requires outdated full
CSI samples and thus is more robust to signaling latency in practical
wireless networks. Third, the quadratic optimization problem at each
iteration can be efficiently solved by the Lagrange dual method, which
has very low complexity. Finally, SSCA-THP provides a systematic solution
for the design of THP, which opens the door to solving the more difficult
THP optimization problems that occur in practice. 
\item \textbf{Specific SSCA-THP algorithm design for important applications:}
We apply SSCA-THP to solve several important THP optimization problems
in massive MIMO. Simulations verify the advantages of the proposed
algorithmic framework over existing baseline solutions.
\end{itemize}

The rest of the paper is organized as follows. In Section \ref{sec:System-Model},
we present the system model for the massive MIMO downlink with THP,
various implementation methods for the RF precoder, and the general
THP optimization formulation. The SSCA-THP algorithm and the convergence
analysis are presented in Section \ref{sec:Stochastic-Successive-Convex}
and \ref{sec:Convergence-Analysis}, respectively. Section \ref{sec:Applications-and-Numerical}
applies SSCA-THP to solve several important THP optimization problems.
Finally, the conclusion is given in Section \ref{sec:Conlusion}.

\textit{Notation}\emph{s}: $\left|\mathcal{S}\right|$ denotes the
cardinality of a set $\mathcal{S}$. $\textrm{Diag}\left(\boldsymbol{a}\right)$
represents a diagonal matrix whose diagonal elements form the vector
$\boldsymbol{a}$. For a matrix $\boldsymbol{M}$, $\textrm{Diag}\left(\boldsymbol{M}\right)$
denotes a vector consisting of the diagonal elements of $\boldsymbol{M}$
and $\left[\boldsymbol{M}\right]_{i,j}$ denotes the $\left(i,j\right)$-th
element of $\boldsymbol{M}$. Let $\boldsymbol{M}=\textrm{BlockDiag}\left(\boldsymbol{M}_{1},\boldsymbol{M}_{2},...,\boldsymbol{M}_{n}\right)$
denote a block diagonal matrix with the $i$-th block given by $\boldsymbol{M}_{i}$,
and $\mathfrak{S}\left[\boldsymbol{M}\right]\triangleq\boldsymbol{M}+\boldsymbol{M}^{H}$.
Let $\circ$ denote the Hadamard product and $\mathfrak{R}[\boldsymbol{M}]$
denote the real part of a complex matrix $\boldsymbol{M}$.

\begin{table}
\begin{centering}
{\footnotesize{}}%
\begin{tabular}{|l|l|}
\hline 
{\small{}DPS } & {\small{}Dynamic phase shifter}\tabularnewline
\hline 
{\small{}FHP} & {\small{}Fast-timescale hybrid precoding}\tabularnewline
\hline 
{\small{}MWTM} & {\small{}Minimum weighted throughput maximization}\tabularnewline
\hline 
{\small{}MM} & {\small{}Majorization-minimization}\tabularnewline
\hline 
{\small{}SSCA} & {\small{}Stochastic successive convex approximation}\tabularnewline
\hline 
{\small{}SAA} & {\small{}Sample average approximation}\tabularnewline
\hline 
{\small{}THP} & {\small{}Two-timescale hybrid precoding}\tabularnewline
\hline 
\end{tabular}
\par\end{centering}{\footnotesize \par}
{\small{}\caption{\label{tab:Cputime-1}{\small{}List of abbreviations.}}
}{\small \par}
\end{table}

\section{System Model and Problem Formulation\label{sec:System-Model}}

\subsection{Multi-user Massive MIMO Downlink with THP}

\begin{figure}
\begin{centering}
\includegraphics[width=75mm]{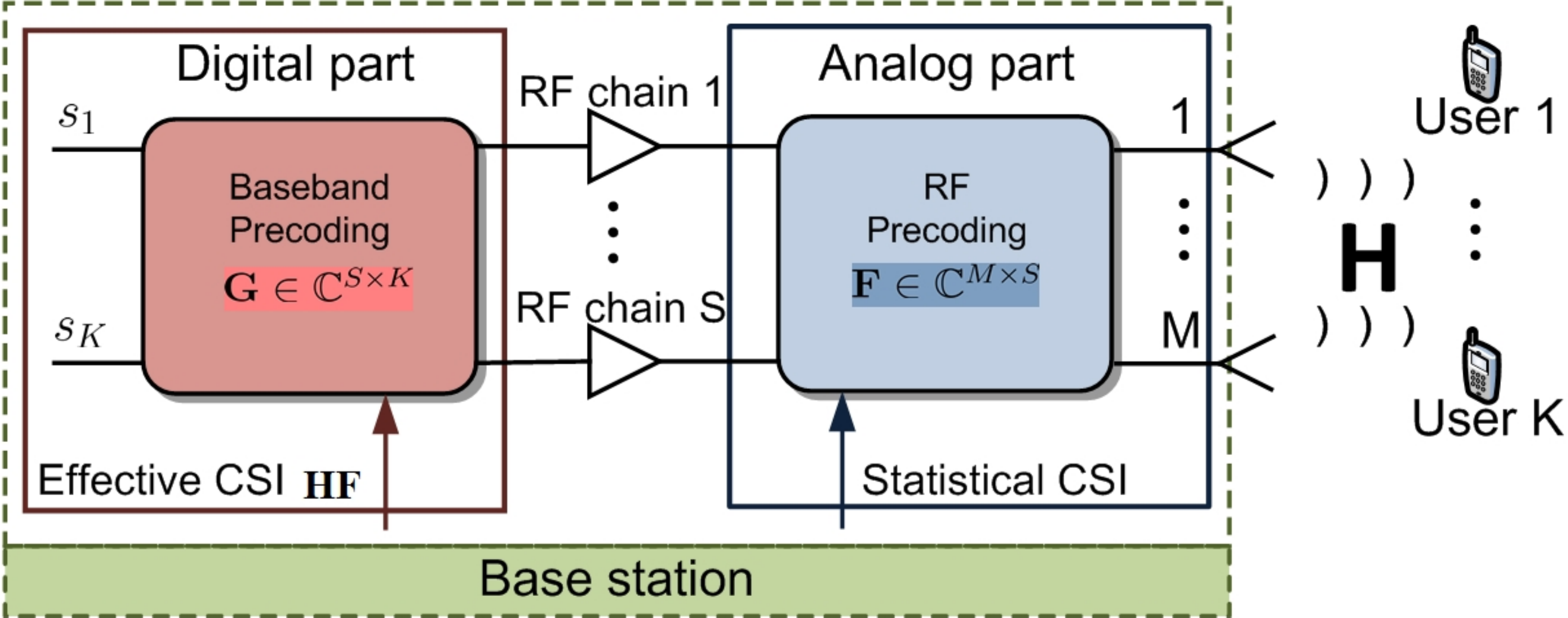}
\par\end{centering}
\caption{\label{fig:sysmodel}{\small{}Massive MIMO downlink with THP}}
\end{figure}

Consider a multi-user massive MIMO downlink system with one BS serving
$K$ single-antenna users, as illustrated in Fig. \ref{fig:sysmodel}.
For clarity, we focus on a narrowband system with flat block fading
channel, but the proposed algorithm can be easily modified to cover
the wideband system as well. The BS is equipped with $M$ antennas
and $S$ transmit RF chains, where $K\leq S\ll M$. Hybrid precoding
is employed to support simultaneous transmissions to the $K$ users
with limited RF chains at the BS. In this case, the transmit signal
vector for user $k$ is given by $\boldsymbol{F}\boldsymbol{g}_{k}s_{k},$
where $\boldsymbol{F}\in\mathbb{C}^{M\times S}$ is the RF precoder,
and $\boldsymbol{g}_{k}\in\mathbb{C}^{S\times1}$ and $s_{k}\sim\mathcal{CN}\left(0,1\right)$
are the baseband precoding vector and the data symbol for user $k$,
respectively. The RF precoder $\boldsymbol{F}$ is usually implemented
using an RF phase shifting network \cite{Zhang_TSP05_RFshifter}.
Hence, all elements of $\boldsymbol{F}$ have equal magnitude, i.e.,
$\left[\boldsymbol{F}\right]_{i,j}=\frac{1}{\sqrt{M}}e^{\sqrt{-1}\theta_{i,j}}$,
where $\theta_{i,j}$ is the phase of the $\left(i,j\right)$-th element
of $\boldsymbol{F}$. Under hybrid precoding, the received signal
for user $k$ is given by
\begin{equation}
y_{k}=\sqrt{p_{k}}\boldsymbol{h}_{k}^{H}\boldsymbol{F}\boldsymbol{g}_{k}s_{k}+\boldsymbol{h}_{k}^{H}\sum_{i\neq k}\sqrt{p_{i}}\boldsymbol{F}\boldsymbol{g}_{i}s_{i}+z_{k},\label{eq:Rxsignal}
\end{equation}
where $\boldsymbol{h}_{k}\in\mathbb{C}^{M}$ is the channel of user
$k$, $p_{k}$ is the transmit power allocated to user $k$, and $z_{k}\sim\mathcal{CN}\left(0,1\right)$
is the additive white Gaussian noise (AWGN). 

\begin{figure}
\begin{centering}
\includegraphics[width=80mm]{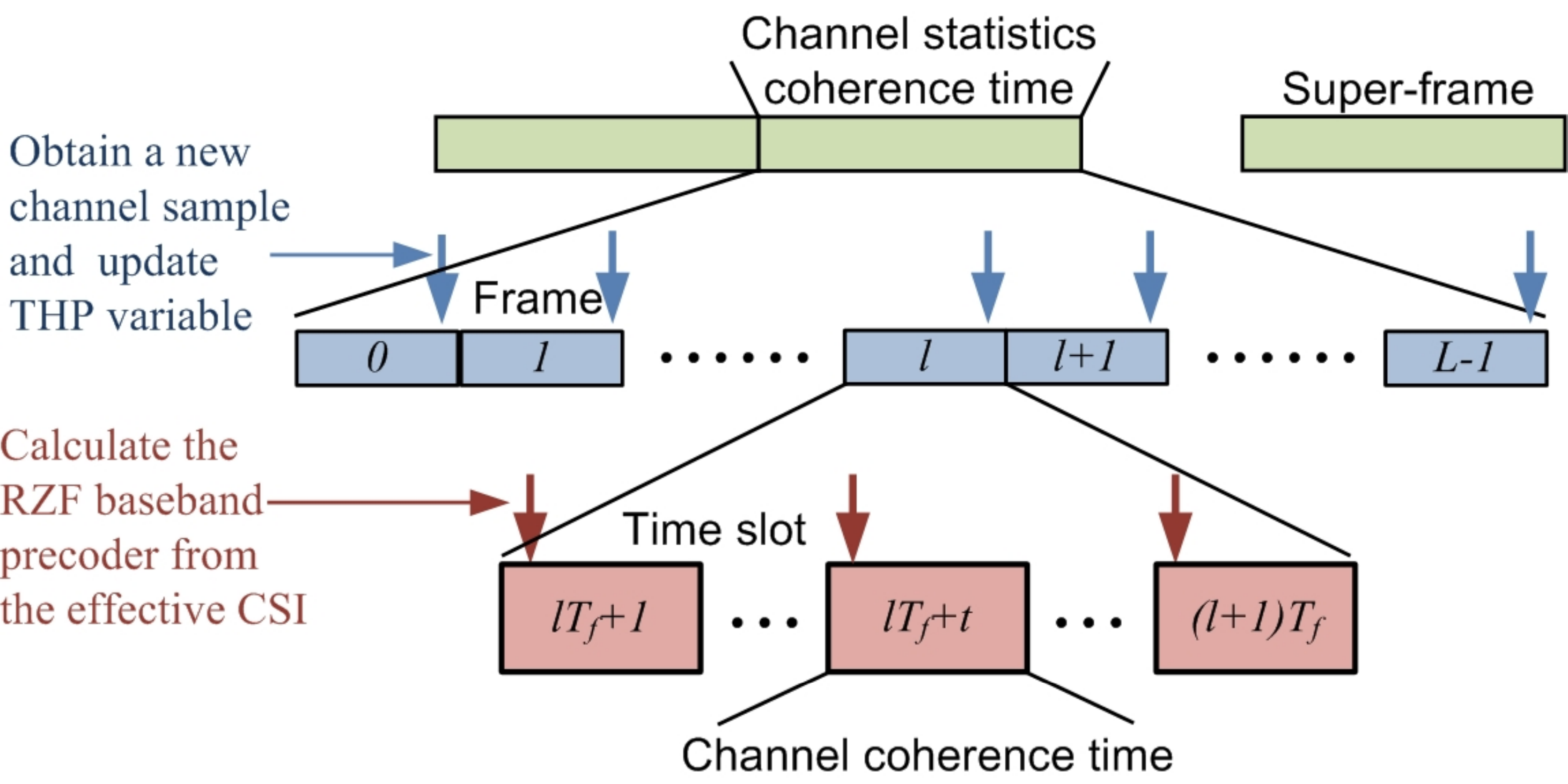}
\par\end{centering}
\caption{\label{fig:timelineAlg}{\small{}Timeline (frame structure) of SSCA-THP
algorithm}}
\end{figure}

In this paper, we consider a THP scheme, whose time line is illustrated
in Fig. \ref{fig:timelineAlg}. The time domain is divided into super-frames.
Each super-frame is further divided into $L$ frames, and each frame
consists of $T_{f}$ time slots. The channel statistics (distribution)
are assumed to be constant within each super-frame, and the channel
state $\boldsymbol{H}$ is assumed to be constant within each time
slot. We assume that the BS can obtain the real-time effective CSI
$\boldsymbol{H}\boldsymbol{F}$ at each time slot, and one (possibly
outdated) channel sample $\boldsymbol{H}$ at each frame. In our design,
the BS is not required to have explicit knowledge of the channel statistics.
By observing one channel sample at each frame, the proposed algorithm
can automatically learn the channel statistics (in an implicit way).
Specifically, the RF precoder $\mathbf{F}$ is only updated once per
frame based on the (possibly outdated) channel sample to achieve massive
MIMO array gain with reduced implementation cost. On the other hand,
the baseband precoder $\boldsymbol{G}=\left[\boldsymbol{g}_{1},...,\boldsymbol{g}_{K}\right]$
is adaptive to the real-time effective CSI $\boldsymbol{H}\boldsymbol{F}\in\mathbb{C}^{K\times S}$
to achieve the spatial multiplexing gain, where $\boldsymbol{H}=\left[\boldsymbol{h}_{1},...,\boldsymbol{h}_{K}\right]^{H}\in\mathbb{C}^{K\times M}$
is the composite downlink channel. We consider a regularized zero-forcing
(RZF) baseband precoder \cite{Peel_TOC05_RCI}:
\begin{equation}
\boldsymbol{G}=\boldsymbol{F}^{H}\boldsymbol{H}^{H}\left(\boldsymbol{H}\boldsymbol{F}\boldsymbol{F}^{H}\boldsymbol{H}^{H}+\alpha\boldsymbol{I}\right)^{-1}\boldsymbol{\Lambda}^{1/2},\label{eq:Geq}
\end{equation}
where $\alpha$ is the regularization factor, $\boldsymbol{\Lambda}=\textrm{Diag}\left(\left[\left\Vert \overline{\boldsymbol{g}}_{1}\right\Vert ^{-1},...,\left\Vert \overline{\boldsymbol{g}}_{K}\right\Vert ^{-1}\right]\right)$
is used to normalize the precoding vectors $\boldsymbol{F}\boldsymbol{g}_{k}$'s,
and $\overline{\boldsymbol{g}}_{k}$ is the $k$-th column of $\overline{\boldsymbol{G}}\triangleq\boldsymbol{F}\boldsymbol{F}^{H}\boldsymbol{H}^{H}\left(\boldsymbol{H}\boldsymbol{F}\boldsymbol{F}^{H}\boldsymbol{H}^{H}+\alpha\boldsymbol{I}\right)^{-1}$.

Although the baseband precoder $\boldsymbol{G}$ is adaptive to the
instantaneous effective CSI, as in (\ref{eq:Geq}), we assume that
the regularization factor $\alpha$ and the power allocation $\boldsymbol{p}=\left[p_{1},...,p_{K}\right]^{T}$
are adaptive to the channel statistics only. This is because in the
massive MIMO regime, the system tends to behave like a deterministic
system and thus the gain of adapting the power allocation and regularization
factor according to the instantaneous CSI is small \cite{Wagner_TIT12s_LargeMIMO,Liu_TSP14_CRANAS}.
Similar assumptions have also been made in \cite{Caire_TIT13_JSDM,Liu_TSP14_RFprecoding,Park_TSP17_THP}
to achieve a good compromise between performance and complexity.

\subsection{Various Implementation Methods for RF Precoder\label{subsec:Various-Implementation-Methods}}

Various implementation methods for the RF precoder $\boldsymbol{F}$
have been proposed in the literature to achieve different tradeoffs
between the performance, complexity and power consumption. Basically,
there are two major RF precoding structures: the fully-connected structure
and the partially-connected structure. For each structure, there are
two common methods to dynamically adjust the RF precoder. In this
subsection, we shall discuss the existing typical implementations
for the RF precoder and their pros and cons. 

One major challenge for the optimization of the RF precoder is that
it has a discrete implementation constraint; e.g., the phase shifter
cannot take a continuous value in practice and has to be quantized
into discrete values. In this subsection, we will also discuss two
techniques to convert the discrete RF precoder into continuous variables
to make the optimization of the RF precoder tractable.

\subsubsection{Fully-connected RF Precoding Structure}

\begin{figure}
\begin{centering}
\includegraphics[width=80mm]{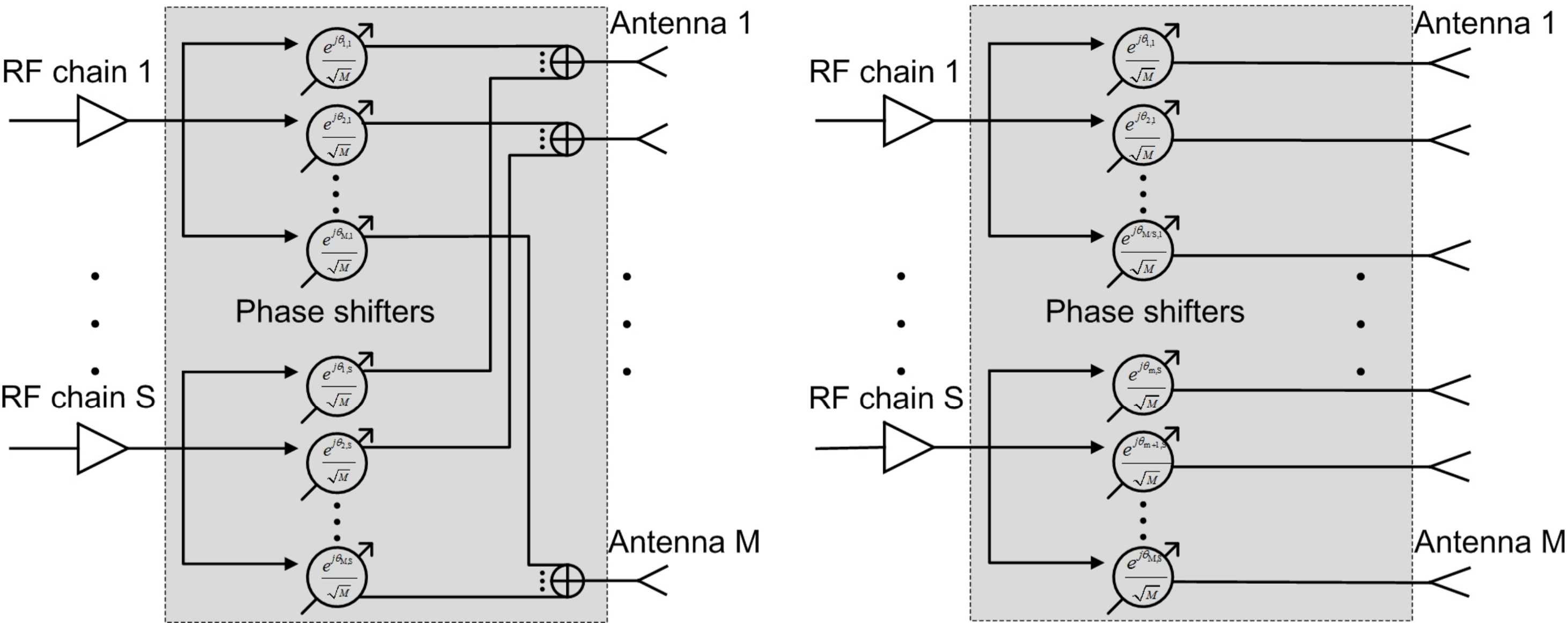}
\par\end{centering}
\caption{\label{fig:fullsubstruc}{\small{}Fully-connected and partially-connected
structures in RF precoder}}
\end{figure}

In this case, each RF chain is connected to every antenna through
phase shifters and RF adders. To be more specific, if a BS has $M$
antennas and $S$ RF chains, $MS$ phase shifters and $M$ RF adders
will be required to implement the RF precoder, as illustrated in Fig.
\ref{fig:fullsubstruc}-(a). There are two methods to dynamically
adjust the RF precoder, as elaborated below. 

\textbf{DPS-based RF Precoder:} In this method, each phase shifter
is quantized using $B$ bits. Then each phase shifter $\theta_{i,j}$
can take a value from the discrete set $\left\{ 0,\frac{2\pi}{2^{B}},...,\frac{2\pi\left(2^{B}-1\right)}{2^{B}}\right\} $.
Since the phase of each element of the RF precoder can be quantized
and adjusted individually, the DPS-based RF precoder can achieve a
good performance. However, the requirement of a large number of dynamic
phase shifters and RF adders increases the hardware cost and power
consumption. 

The DPS-based RF precoder can be represented by a phase vector $\boldsymbol{\theta}\in\mathbb{R}^{MS}$
whose $\left(\left(j-1\right)M+i\right)$-th element is $\theta_{i,j}$,
and the optimization of the DPS-based RF precoder is a discrete optimization
problem due to the discrete constraint on $\theta_{i,j}$. To make
the problem tractable, we first ignore the discrete constraint and
treat $\theta_{i,j}$ as a continuous variable. After finding the
optimized phase vector $\boldsymbol{\theta}^{*}$, we project it onto
the feasible set to obtain the final phase vector $\hat{\boldsymbol{\theta}}$:
\begin{align}
\hat{\theta}_{i,j} & =\underset{\theta\in\left\{ 0,...,2\pi\left(2^{B}-1\right)/2^{B}\right\} }{\text{argmin}}\left|\left(\theta_{i,j}^{*}\textrm{ mod }2\pi\right)-\theta\right|,\label{eq:Projtheta}
\end{align}
$\forall i,j$, where $\textrm{mod}$ denotes the modulo operation.
With only $B=3$ bits, the performance loss due to the quantization
effect is already small \cite{Yu_JSTSP2016_HP}.

\textbf{Codebook-based RF Precoder:} In this case, each column of
the RF precoder can only be selected from a finite-size codebook $\mathcal{F}=\left\{ \boldsymbol{c}_{1},...,\boldsymbol{c}_{N}\right\} $
with $\left|\mathcal{F}\right|=N$ code vectors. The RF precoding
codebook $\mathcal{F}$ is usually chosen to be the array response
vectors at the BS. For example, when a uniform linear array (ULA)
is used, the array response vectors form a DFT matrix. The codebook-based
RF precoder can be easily implemented using a static precoder at the
RF domain (using a static phase shifting network) together with an
RF switch \cite{Liu_TSP14_RFprecoding}. However, the performance
of the codebook-based RF precoder is in general worse than that of
the DPS-based RF precoder.

The codebook-based RF precoder can be represented by a selection matrix
as $\boldsymbol{F}=\boldsymbol{C}\boldsymbol{D}$, where $\boldsymbol{C}=\left[\boldsymbol{c}_{1},...,\boldsymbol{c}_{N}\right]\in\mathbb{C}^{M\times N}$
is the codebook matrix, and $\boldsymbol{D}\triangleq\textrm{Diag}\left(\boldsymbol{d}\right)$
with $\boldsymbol{d}=\left[d_{1},...,d_{N}\right]^{T}\in\left\{ 0,1\right\} ^{N}$
and $\sum_{i=1}^{N}d_{i}=S$ is a selection matrix. The optimization
of the codebook-based RF precoder is an integer optimization problem
due to the constraint $d_{i}\in\left\{ 0,1\right\} ,\forall i=1,...,N$.
Note that the constraints $\sum_{i=1}^{N}d_{i}=S$ and $d_{i}\in\left\{ 0,1\right\} ,\forall i$
are equivalent to the following constraints:
\begin{equation}
\sum_{i=1}^{N}d_{i}=S,\:d_{i}\in\left[0,1\right],\:\left\Vert \boldsymbol{d}\right\Vert _{0}\leq S.\label{eq:sparsed}
\end{equation}
To make the problem tractable, we approximate the $l_{0}$-norm $\left\Vert \boldsymbol{d}\right\Vert _{0}$
using a smooth function as \cite{Wipf_JSTSP10_rwl1norm}
\begin{equation}
\left\Vert \boldsymbol{d}\right\Vert _{0}\approx\sum_{i=1}^{N}\log\left(1+\frac{d_{i}}{\epsilon}\right)/\log\left(1+\frac{1}{\epsilon}\right),\:\boldsymbol{d}\in\left[0,1\right]^{N},\label{eq:SPd}
\end{equation}
where the smooth parameter $\epsilon>0$ can be used to control the
approximation error. A smaller $\epsilon$ leads to smaller approximation
error but a less smooth function. Then we can replace the constraint
$\left\Vert \boldsymbol{d}\right\Vert _{0}\leq S$ with a non-convex
\textit{sparse constraint} 
\begin{equation}
\sum_{i=1}^{N}\log\left(1+\frac{d_{i}}{\epsilon}\right)/\log\left(1+\frac{1}{\epsilon}\right)\leq S,\label{eq:sparsecond}
\end{equation}
which can be handled using the SSCA-THP algorithm. Note that the sparse
constraint (\ref{eq:sparsecond}) has been widely used in compressive
sensing to impose sparsity on sparse signals \cite{Wipf_JSTSP10_rwl1norm}.
After finding the optimized selection vector $\boldsymbol{d}^{*}$,
we project it onto the feasible set to obtain the final solution $\hat{\boldsymbol{d}}$:
\[
\hat{\boldsymbol{d}}=\underset{\boldsymbol{d}}{\textrm{argmin}}\left\Vert \boldsymbol{d}-\boldsymbol{d}^{*}\right\Vert ,\textrm{ s.t.}\:\boldsymbol{d}\in\left\{ 0,1\right\} ^{N},\left\Vert \boldsymbol{d}\right\Vert _{0}=S.
\]
Specifically, $\hat{\boldsymbol{d}}$ is a binary vector whose $S$
non-zero elements are located at the $S$ largest elements of $\boldsymbol{d}^{*}$.
Thanks to the sparse constraint on $\boldsymbol{d}^{*}$, $\boldsymbol{d}^{*}$
is usually close to a binary vector with $S$ non-zero elements, and
thus the performance loss due to the above projection is very small.

\subsubsection{Partially-connected RF Precoding Structure}

In this case, each RF chain is connected to a sub-array of antennas
via phase shifters and each antenna is only connected to a single
RF chain \cite{Molisch_Csurvey2016_HBF}, as illustrated in Fig. \ref{fig:fullsubstruc}-(b).
Such a partially-connected structure presents a block diagonal form
as $\boldsymbol{F}=\textrm{BlockDiag}\left(\boldsymbol{f}_{1},\boldsymbol{f}_{2},...,\boldsymbol{f}_{S}\right)$,
where $\boldsymbol{f}_{s}\in\mathbb{C}^{M/S}$ is the $s$-th precoding
vector corresponding to the $s$-th RF chain. Note that we have assumed
that $M$ is divisible by $S$ for easy illustration. The partially-connected
structure only requires a total number of $M$ phase shifters (instead
of $MS$ phase shifters in the fully-connected case). As a result,
it has much lower power consumption and hardware complexity compared
to the fully-connected case. However, the performance of the partially-connected
RF precoder is compromised. 

The DPS-based or codebook-based method can be used to adjust the partially-connected
RF precoder. In the DPS-based method, the partially-connected RF precoder
can be represented by a phase vector $\boldsymbol{\theta}\in\mathbb{R}^{M}$
whose $\left(\left(j-1\right)M/S+i\right)$-th element is $\theta_{i,j}$
for $i=\left(j-1\right)M/S+1,...,jM/S$; $j=1,...,S$. Similarly,
to make the problem tractable, we first ignore the discrete constraint
on $\theta_{i,j}$ and then project the resulting solution onto the
discrete set using (\ref{eq:Projtheta}).

In the codebook-based method, the $s$-th RF precoding vector $\boldsymbol{f}_{s}$
is selected from a finite-size codebook $\mathcal{F}_{s}=\left\{ \mathbf{c}_{1},...,\mathbf{c}_{N}\right\} $
with $\left|\mathcal{F}_{s}\right|=N$ code vectors. Let $\boldsymbol{C}_{s}=\left[\boldsymbol{c}_{s,1},...,\boldsymbol{c}_{s,N}\right]\in\mathbb{C}^{\frac{M}{S}\times N}$
denote the codebook matrix for $\boldsymbol{f}_{s}$. Then the $s$-th
RF precoding vector $\boldsymbol{f}_{s}$ can be represented as $\boldsymbol{f}_{s}=\boldsymbol{C}_{s}\boldsymbol{D}_{s}$,
where $\boldsymbol{D}_{s}\triangleq\textrm{Diag}\left(\boldsymbol{d}_{s}\right)$
with $\boldsymbol{d}_{s}=\left[d_{s,1},...,d_{s,N}\right]^{T}\in\left\{ 0,1\right\} ^{N}$
and $\sum_{i=1}^{N}d_{s,i}=1$ is a selection matrix. For convenience,
let $\boldsymbol{d}=\left[\boldsymbol{d}_{1},...,\boldsymbol{d}_{S}\right]$
denote the overall selection vector. Similarly, we can use the ``sparse''
technique to relax the integer constraints $d_{s,i}\in\left\{ 0,1\right\} ,\forall s,i$
to non-convex smooth constraints $\sum_{i=1}^{N}d_{s,i}=1,\:d_{s,i}\in\left[0,1\right],\:\sum_{i=1}^{N}\log\left(1+\frac{d_{s,i}}{\epsilon}\right)/\log\left(1+\frac{1}{\epsilon}\right)\leq1.$

\subsection{Achievable Data Rate}

Under different implementation methods, the RF precoder $\boldsymbol{F}$
is represented by different parameters with different dimensions.
For convenience, we use $\boldsymbol{\phi}$ as a unified notation
to denote the \textit{RF precoding parameter}. For example, in the
DPS-based method, $\boldsymbol{\phi}=\boldsymbol{\theta}$ and its
dimension is $MS$ and $M$ for the fully-connected and partially-connected
structures, respectively. In the codebook-based method, $\boldsymbol{\phi}=\boldsymbol{d}$
and its dimension is $N$ and $NS$ for the fully-connected and partially-connected
structures, respectively. 

For given RF precoding parameter $\boldsymbol{\phi},$ power allocation
$\boldsymbol{p}$, the RZF regularization factor $\alpha$ in (\ref{eq:Geq})
and channel realization $\boldsymbol{H}$, the instantaneous achievable
data rate of user $k$ is
\[
r_{k}\left(\boldsymbol{\phi},\boldsymbol{p},\alpha;\boldsymbol{H}\right)=\log\left(1+\frac{p_{k}\left|\boldsymbol{h}_{k}^{H}\boldsymbol{F}\boldsymbol{g}_{k}\right|^{2}}{\sum_{i\neq k}p_{i}\left|\boldsymbol{h}_{k}^{H}\boldsymbol{F}\boldsymbol{g}_{i}\right|^{2}+1}\right).
\]
Note that $\boldsymbol{F}$ is a function of $\boldsymbol{\phi}$
and $\boldsymbol{G}$ is a function of $\boldsymbol{\phi},\boldsymbol{p},\alpha$
and $\boldsymbol{H}$. Therefore, we explicitly express $r_{k}$ as
a function of $\boldsymbol{\phi},\boldsymbol{p},\alpha$ which depends
on the random channel state $\boldsymbol{H}$. The average data rate
of user $k$ is
\[
\overline{r}_{k}\left(\boldsymbol{\phi},\boldsymbol{p},\alpha\right)=\mathbb{E}\left[r_{k}\left(\boldsymbol{\phi},\boldsymbol{p},\alpha;\boldsymbol{H}\right)\right].
\]
For convenience, define $\overline{\boldsymbol{r}}\left(\boldsymbol{\phi},\boldsymbol{p},\alpha\right)\triangleq[\overline{r}_{1}\left(\boldsymbol{\phi},\boldsymbol{p},\alpha\right),...,\overline{r}_{K}\left(\boldsymbol{\phi},\boldsymbol{p},\alpha\right)]^{T}$
as the average data rate vector.

\subsection{THP Optimization Formulation\label{subsec:A-General-Optimization}}

Consider the following general optimization formulation for the design
of THP:
\begin{align}
 & \min_{\boldsymbol{x}\in\mathcal{X}}\:f_{0}\left(\boldsymbol{x}\right)\triangleq h_{0}\left(\overline{\boldsymbol{r}}\left(\boldsymbol{\phi},\boldsymbol{p},\alpha\right),\boldsymbol{x}\right)\label{eq:mainP}\\
\textrm{s.t.} & f_{i}\left(\boldsymbol{x}\right)\triangleq h_{i}\left(\overline{\boldsymbol{r}}\left(\boldsymbol{\phi},\boldsymbol{p},\alpha\right),\boldsymbol{x}\right)\leq0,i=1,....,m,\nonumber 
\end{align}
where $\boldsymbol{x}=\left[\boldsymbol{\phi}^{T},\boldsymbol{p}^{T},\alpha,\boldsymbol{\beta}^{T}\right]^{T}$
is called the \textit{THP variable}, $\boldsymbol{\beta}\in\mathbb{R}^{n_{\beta}}$,
with dimension $n_{\beta}$, is an additional optimization variable
that does not affect the average data rate vector $\overline{\boldsymbol{r}}$,
and $\mathcal{X}\subseteq\mathbb{R}^{n}$ is a convex set. The motivation
for introducing an additional optimization variable $\boldsymbol{\beta}$
is to provide extra flexibility in the formulation to cover more application
scenarios, as will be shown in Example 4 at the end of this subsection.
Note that both the dimension $n$ and the set $\mathcal{X}$ depend
on the implementation method for the RF precoder. For example, for
the codebook-based RF precoder, we have
\begin{equation}
\mathcal{X}=\left\{ \boldsymbol{x}:\boldsymbol{d}\in\left[0,1\right]^{N};\boldsymbol{p}\in\left[0,\tilde{p}\right]^{K};\alpha>\tilde{\alpha};\boldsymbol{\beta}\in\mathcal{B}\right\} ,\label{eq:CodebookX}
\end{equation}
where $\tilde{p}>0$ is used to ensure that the transmit power is
bounded, $\tilde{\alpha}>0$ is a small number to ensure that the
calculation of the matrix inverse in (\ref{eq:Geq}) is always numerically
stable, $\mathcal{B}=\left\{ \boldsymbol{\beta}:\:\beta_{i}\in\mathcal{B}_{i},i=1,...,n_{\beta}\right\} $,
and $\mathcal{B}_{i}$ is a convex region in $\mathbb{R}$. Note that,
without loss of generality, all the coupled constraints on $\boldsymbol{x}$,
such as the sparse constraint (\ref{eq:sparsecond}) on the codebook-based
RF precoding parameter $\boldsymbol{d}$, are included in the explicit
constraints $f_{i}\left(\boldsymbol{x}\right)\leq0,i=1,....,m$. As
a result, $\mathcal{X}$ has a decoupled form: $\mathcal{X}=\left\{ \boldsymbol{x}:\:x_{i}\in\mathcal{X}_{i},i=1,...,n\right\} $,
where $\mathcal{X}_{i}$ is a convex region in $\mathbb{R}$. We assume
that the functions $h_{i}\left(\overline{\boldsymbol{r}},\boldsymbol{x}\right),i=0,...,m$
are continuously differentiable (and possibly non-convex) functions
of $\left(\overline{\boldsymbol{r}},\boldsymbol{x}\right)$. 

Problem (\ref{eq:mainP}) embraces many applications as special cases.
In the following, we give some important examples. 
\begin{example}
[Sum throughput maximization \cite{Liu_TSP2016_CSImassive}]\label{exa:sum}The
sum throughput maximization problem is formulated as:
\begin{equation}
\max_{\boldsymbol{x}\in\mathcal{X}}\:\sum_{k=1}^{K}\overline{r}_{k}\left(\boldsymbol{\phi},\boldsymbol{p},\alpha\right),\textrm{ s.t. }\sum_{k=1}^{K}p_{k}\leq P,\label{eq:sum}
\end{equation}
where $P$ is the total power constraint at the BS. This is an instance
of Problem (\ref{eq:mainP}) with $h_{0}\left(\overline{\boldsymbol{r}},\boldsymbol{x}\right)=-\sum_{k=1}^{K}\overline{r}_{k}$,
$h_{1}\left(\overline{\boldsymbol{r}},\boldsymbol{x}\right)=\sum_{k=1}^{K}p_{k}-P$
and $\mathcal{B}=\emptyset$ (i.e., there is no additional variable
$\boldsymbol{\beta}$).
\end{example}

\begin{example}
[PFS \cite{Kelly_OPR98_PFS}]\label{exa:PFS}PFS is a widely used
utility function in wireless resource optimization. The PFS utility
maximization problem is formulated as:
\begin{equation}
\max_{\boldsymbol{x}\in\mathcal{X}}\:\sum_{k=1}^{K}\textrm{log}\left(\varepsilon+\overline{r}_{k}\left(\boldsymbol{\phi},\boldsymbol{p},\alpha\right)\right),\textrm{ s.t. }\sum_{k=1}^{K}p_{k}\leq P,\label{eq:PFS}
\end{equation}
where $\varepsilon>0$ is a small number used to avoid the singularity
at $\overline{r}_{k}=0$ \cite{Liu_TSP13_HierarchicalmassiveMIMO}.
This is an instance of Problem (\ref{eq:mainP}) with $h_{0}\left(\overline{\boldsymbol{r}},\boldsymbol{x}\right)=-\sum_{k=1}^{K}\textrm{log}\left(\varepsilon+\overline{r}_{k}\right)$,
$h_{1}\left(\overline{\boldsymbol{r}},\boldsymbol{x}\right)=\sum_{k=1}^{K}p_{k}-P$
and $\mathcal{B}=\emptyset$.
\end{example}

\begin{example}
[Power minimization with individual QoS requirements \cite{Larsson_TWC2016_PowminMMIMO}]\label{exa:Portfolio-optimization-1}In
this example, the THP variable $\boldsymbol{x}$ is designed to minimize
the average transmit power subject to individual QoS requirements
as follows:
\begin{equation}
\min_{\boldsymbol{x}\in\mathcal{X}}\:\sum_{k=1}^{K}p_{k},\textrm{ s.t. }\overline{r}_{k}\left(\boldsymbol{\phi},\boldsymbol{p},\alpha\right)\geq\gamma_{k},\forall k,\label{eq:IQoS}
\end{equation}
where each user has an individual QoS requirement in terms of the
average data rate constraint $\overline{r}_{k}\left(\boldsymbol{\phi},\boldsymbol{p},\alpha\right)\geq\gamma_{k}$,
and the constant $\gamma_{k}\geq0$ is the target rate for user $k$.
This is an instance of Problem (\ref{eq:mainP}) with $h_{0}\left(\overline{\boldsymbol{r}},\boldsymbol{x}\right)=\sum_{k=1}^{K}p_{k}$,
$h_{k}\left(\overline{\boldsymbol{r}},\boldsymbol{x}\right)=\gamma_{k}-\overline{r}_{k},\forall k$
and $\mathcal{B}=\emptyset$.
\end{example}

\begin{example}
[MWTM \cite{Liu_TSP14_RFprecoding}]\label{exa:Portfolio-optimization}In
this example, the THP variable $\boldsymbol{x}$ is designed to maximize
the minimum (weighted) average data rate of users as follows:
\begin{equation}
\max_{\boldsymbol{x}\in\mathcal{X}}\:\min_{k}\frac{1}{w_{k}}\overline{r}_{k}\left(\boldsymbol{\phi},\boldsymbol{p},\alpha\right),\textrm{ s.t. }\sum_{k=1}^{K}p_{k}\leq P,\label{eq:MWTMorh}
\end{equation}
where $w_{k}>0$ is the weight for user $k$, which can be used to
provide a differential QoS for different users. (\ref{eq:MWTMorh})
is not an instance of Problem (\ref{eq:mainP}) because the objective
function is non-smooth. However, by introducing an auxiliary variable
$\beta$, we can convert (\ref{eq:MWTMorh}) to an instance of Problem
(\ref{eq:mainP}) as:
\begin{equation}
\max_{\boldsymbol{x}\in\mathcal{X}}\:\beta,\textrm{ s.t. }\overline{r}_{k}\left(\boldsymbol{\phi},\boldsymbol{p},\alpha\right)\geq w_{k}\beta,\forall k;\sum_{k=1}^{K}p_{k}\leq P,\label{eq:MWTM}
\end{equation}
with $h_{0}\left(\overline{\boldsymbol{r}},\boldsymbol{x}\right)=-\beta$,
$h_{k}\left(\overline{\boldsymbol{r}},\boldsymbol{x}\right)=w_{k}\beta-\overline{r}_{k},\forall k$,
$h_{K+1}\left(\overline{\boldsymbol{r}},\boldsymbol{x}\right)=\sum_{k=1}^{K}p_{k}-P$
and $\mathcal{B}=\left\{ \beta:\:\beta\geq0\right\} $.
\end{example}

There are several challenges to solve Problem (\ref{eq:mainP}). First,
the average data rates $\overline{r}_{k}\left(\boldsymbol{\phi},\boldsymbol{p},\alpha\right)$'s
are neither convex nor concave, and have no closed-form expressions.
Moreover, the presence of stochastic non-convex constraints further
complicates Problem (\ref{eq:mainP}). In the next section, we shall
propose an efficient algorithm based on the SSCA method, called \textit{SSCA-THP},
to find a stationary point of Problem (\ref{eq:mainP}). 

\section{Stochastic Successive Convex Approximation for THP Optimization\label{sec:Stochastic-Successive-Convex}}

\subsection{The SSCA-THP Algorithm\label{subsec:The-SSCA-THP-Algorithm}}

At each iteration, the THP variable $\boldsymbol{x}$ is updated by
solving a quadratic optimization problem obtained by replacing the
objective and constraint functions $f_{i}\left(\boldsymbol{x}\right),i=0,...,m$
with their quadratic surrogate functions $\bar{f}_{i}^{l}\left(\boldsymbol{x}\right),i=0,...,m$. 

Algorithm \ref{alg1} summarizes the key steps of the proposed SSCA-THP
algorithm. Specifically, at iteration $l$, a new realization of the
random channel state $\boldsymbol{H}^{l}$ is obtained in Step 1 and
the surrogate functions $\bar{f}_{i}^{l}\left(\boldsymbol{x}\right),\forall i$
are updated based on $\boldsymbol{H}^{l}$ and the current iterate
$\boldsymbol{x}^{l}$ as
\begin{align}
\bar{f}_{i}^{l}\left(\boldsymbol{x}\right) & =h_{i}\left(\hat{\boldsymbol{r}}^{l},\boldsymbol{x}^{l}\right)+\left(\mathbf{u}_{i}^{l}\right)^{T}\left(\boldsymbol{x}-\boldsymbol{x}^{l}\right)+\tau_{i}\left\Vert \boldsymbol{x}-\boldsymbol{x}^{l}\right\Vert ^{2},\label{eq:upsurrgate}
\end{align}
where $\tau_{i}>0$ is a constant; $\hat{\boldsymbol{r}}^{l}=\left[\hat{r}_{1}^{l},...,\hat{r}_{K}^{l}\right]^{T}$,
with $\hat{r}_{k}^{l}=\sum_{j=1}^{l}r_{k}\left(\boldsymbol{\phi}^{l},\boldsymbol{p}^{l},\alpha^{l};\boldsymbol{H}^{j}\right)/l$,
is the sample average approximations for $\overline{r}_{k}\left(\boldsymbol{\phi}^{l},\boldsymbol{p}^{l},\alpha^{l}\right)$;
$\mathbf{u}_{i}^{l}$ is an approximation for the gradient $\nabla f_{i}\left(\boldsymbol{x}^{l}\right)$,
which is updated recursively as
\begin{align*}
\mathbf{u}_{i}^{l} & =\left(1-\rho^{l}\right)\mathbf{u}_{i}^{l-1}+\rho^{l}\hat{\mathbf{u}}_{i}^{l},
\end{align*}
with $\mathbf{u}^{-1}=\boldsymbol{0}$, where $\rho^{l}\in\left(0,1\right]$
is a sequence to be properly chosen and
\begin{align}
\hat{\mathbf{u}}_{i}^{l} & =\mathbf{J}_{r}\left(\boldsymbol{x}^{l};\boldsymbol{H}^{l}\right)\nabla_{\overline{\boldsymbol{r}}}h_{i}\left(\hat{\boldsymbol{r}}^{l},\boldsymbol{x}^{l}\right)+\nabla_{\boldsymbol{x}}h_{i}\left(\hat{\boldsymbol{r}}^{l},\boldsymbol{x}^{l}\right),\label{eq:uhead}
\end{align}
where $\mathbf{J}_{r}\left(\boldsymbol{x}^{l};\boldsymbol{H}^{l}\right)$
is the Jacobian matrix of the instantaneous rate vector $\boldsymbol{r}\left(\boldsymbol{\phi},\boldsymbol{p},\alpha;\boldsymbol{H}\right)\triangleq[r_{1}\left(\boldsymbol{\phi},\boldsymbol{p},\alpha;\boldsymbol{H}\right),...,r_{K}\left(\boldsymbol{\phi},\boldsymbol{p},\alpha;\boldsymbol{H}\right)]^{T}$
and its expression is derived in Appendix \ref{subsec:Jacobian-Matrix-of},
$\nabla_{\overline{\boldsymbol{r}}}h_{i}$ and $\nabla_{\boldsymbol{x}}h_{i}$
are the gradients of $h_{i}$ w.r.t. the average rate vector $\overline{\boldsymbol{r}}$
and the THP variable $\boldsymbol{x}$, respectively. The surrogate
function $\bar{f}_{i}^{l}\left(\boldsymbol{x}\right)$ can be viewed
as a convex approximation of $f_{i}\left(\boldsymbol{x}\right)$ in
a local domain around $\boldsymbol{x}^{l}$. 

In Step 2, the optimal solution $\bar{\boldsymbol{x}}^{l}$ of the
following problem is solved:
\begin{align}
\bar{\boldsymbol{x}}^{l}=\underset{\boldsymbol{x}\in\mathcal{X}}{\text{argmin}}\: & \bar{f}_{0}^{l}\left(\boldsymbol{x}\right)\label{eq:Pitert}\\
s.t.\: & \bar{f}_{i}^{l}\left(\boldsymbol{x}\right)\leq0,i=1,....,m,\nonumber 
\end{align}
which is a convex approximation of (\ref{eq:mainP}). Note that Problem
(\ref{eq:Pitert}) is not necessarily feasible. If Problem (\ref{eq:Pitert})
turns out to be infeasible, the optimal solution $\bar{\boldsymbol{x}}^{l}$
of the following convex problem is solved: 

\begin{align}
\bar{\boldsymbol{x}}^{l}=\underset{\boldsymbol{x}\in\mathcal{X},\nu}{\text{argmin}} & \:\nu\label{eq:Pitert-1}\\
s.t.\: & \bar{f}_{i}^{l}\left(\boldsymbol{x}\right)\leq\nu,i=1,....,m,\nonumber 
\end{align}
which minimizes the constraint functions. 

Given $\bar{\boldsymbol{x}}^{l}$ in one of the above two cases, $\boldsymbol{x}$
is updated in Step 3 according to
\begin{equation}
\boldsymbol{x}^{l+1}=\left(1-\gamma^{l}\right)\boldsymbol{x}^{l}+\gamma^{l}\bar{\boldsymbol{x}}^{l},\label{eq:updatext}
\end{equation}
where $\gamma^{l}\in\left(0,1\right]$ is a sequence to be properly
chosen. Then the above iteration (Steps 1 to 3) is carried out until
convergence.

\begin{algorithm}
\caption{\label{alg1}SSCA-THP Algorithm}

\textbf{Input: }$\left\{ \gamma^{l}\right\} $, $\left\{ \rho^{l}\right\} $.

\textbf{Initialize:} $\boldsymbol{x}^{0}\in\mathcal{X}$; $\mathbf{u}_{i}^{-1}=\boldsymbol{0},\forall i$,
$l=0$.

\textbf{Step 1: }Obtain a channel sample $\boldsymbol{H}^{l}$ within
frame $l$. 

Update\textbf{ }the surrogate functions $\bar{f}_{i}^{l}\left(\boldsymbol{x}\right),\forall i$
using (\ref{eq:upsurrgate}).

\textbf{Step 2: }Solve (\ref{eq:Pitert-1}) to obtain the optimal
solution $\nu^{\circ},\boldsymbol{x}^{\circ}$.

\textbf{If} $\nu^{\circ}\leq0$ (Problem (\ref{eq:Pitert}) is feasible)

Solve (\ref{eq:Pitert}) to obtain $\bar{\boldsymbol{x}}^{l}$. //Objective
update

\textbf{Else }

Let $\bar{\boldsymbol{x}}^{l}=\boldsymbol{x}^{\circ}$. //Feasible
update

\textbf{End if}

\textbf{Step 3: }Update $\boldsymbol{x}^{l+1}$ according to (\ref{eq:updatext}).

\textbf{Step 4: Let} $l=l+1$ and return to Step 1.
\end{algorithm}

\subsection{Efficient Solutions for Quadratic Optimization Subproblems\label{subsec:Efficient-Solutions-for}}

In this subsection, we propose efficient solutions for the quadratic
optimization subproblems in (\ref{eq:Pitert}) and (\ref{eq:Pitert-1})
based on the Lagrange dual method. The reasons for using the Lagrange
dual method are as follows. First, for given Lagrange multipliers
(which are also called dual variables), the problem of minimizing
the Lagrange function has a unique and closed-form solution. Second,
the number of primal variables $\boldsymbol{x}$ is usually much larger
than the number of constraints (dual variables) in the massive MIMO
regime. Therefore, the optimal dual variables can be solved much more
efficiently than directly solving the optimal primal variables.

In the following, we show how to use the Lagrange dual method to solve
the subproblem in (\ref{eq:Pitert}). Since both subproblems have
the same form (i.e., both are strictly convex and quadratic optimization
problems), the solution for (\ref{eq:Pitert-1}) is similar and is
omitted for conciseness. 

The Lagrange function for (\ref{eq:Pitert}) is
\begin{align*}
\mathcal{L}^{l}\left(\boldsymbol{x},\boldsymbol{\lambda}\right) & =\bar{f}_{0}^{l}\left(\boldsymbol{x}\right)+\sum_{i=1}^{m}\lambda_{i}\bar{f}_{i}^{l}\left(\boldsymbol{x}\right)\\
 & =\sum_{i=1}^{n}a\left(\boldsymbol{\lambda}\right)x_{i}^{2}+b_{i}\left(\boldsymbol{\lambda}\right)x_{i}+c\left(\boldsymbol{\lambda}\right),\:\boldsymbol{x}\in\mathcal{X},
\end{align*}
where $\boldsymbol{\lambda}=\left[\lambda_{1},...,\lambda_{m}\right]^{T}\in\mathbb{R}_{+}^{m}$
are the Lagrange multipliers, 
\begin{align*}
a\left(\boldsymbol{\lambda}\right) & =\sum_{j=0}^{m}\lambda_{j}\tau_{j},\\
b_{i}\left(\boldsymbol{\lambda}\right) & =\sum_{j=0}^{m}\lambda_{j}\left(u_{j,i}^{l}-2\tau_{j}x_{i}^{l}\right),\\
c\left(\boldsymbol{\lambda}\right) & =\sum_{j=0}^{m}\lambda_{j}\left(h_{j}\left(\hat{\boldsymbol{r}}^{l},\boldsymbol{x}^{l}\right)-\left(\mathbf{u}_{j}^{l}\right)^{T}\boldsymbol{x}^{l}+\tau_{j}\left\Vert \boldsymbol{x}^{l}\right\Vert ^{2}\right),
\end{align*}
$\lambda_{0}=1$, and $x_{i}^{l}$ and $u_{j,i}^{l}$ are the $i$-th
element of $\boldsymbol{x}^{l}$ and $\mathbf{u}_{j}^{l}$, respectively.
The dual function for (\ref{eq:Pitert}) is
\begin{equation}
g^{l}\left(\boldsymbol{\lambda}\right)=\min_{\boldsymbol{x}\in\mathcal{X}}\mathcal{L}^{l}\left(\boldsymbol{x},\boldsymbol{\lambda}\right).\label{eq:dual}
\end{equation}
And the corresponding dual problem is
\begin{equation}
\max_{\boldsymbol{\lambda}\geq\boldsymbol{0}}\:g^{l}\left(\boldsymbol{\lambda}\right).\label{eq:dualP}
\end{equation}

The minimization problem in (\ref{eq:dual}) can be decomposed into
$n$ independent subproblems as
\[
\min_{x_{i}\in\mathcal{X}_{i}}a\left(\boldsymbol{\lambda}\right)x_{i}^{2}+b_{i}\left(\boldsymbol{\lambda}\right)x_{i},\:i=1,...,n,
\]
which have the following closed-form solutions:
\begin{equation}
x_{i}^{\circ}\left(\boldsymbol{\lambda}\right)=\mathbb{P}_{\mathcal{X}_{i}}\left[-\frac{b_{i}\left(\boldsymbol{\lambda}\right)}{2a\left(\boldsymbol{\lambda}\right)}\right],\forall i,\label{eq:primopt}
\end{equation}
where $\mathbb{P}_{\mathcal{X}_{i}}\left[\cdot\right]$ denotes the
one-dimensional projection on to the convex set $\mathcal{X}_{i}$.
On the other hand, the dual function $g\left(\boldsymbol{\lambda}\right)$
is concave and it can be verified that $\left[\bar{f}_{1}^{l}\left(\boldsymbol{x}^{\circ}\left(\boldsymbol{\lambda}\right)\right),,...,\bar{f}_{m}^{l}\left(\boldsymbol{x}^{\circ}\left(\boldsymbol{\lambda}\right)\right)\right]^{T}$
is a subgradient of $g\left(\boldsymbol{\lambda}\right)$ at $\boldsymbol{\lambda}$.
Hence, the standard subgradient-based methods such as the subgradient
algorithm in \cite{Boyd_03note_Subgradient} or the ellipsoid method
in \cite{Boyd_04Book_Convex_optimization} can be used to solve the
optimal solution $\boldsymbol{\lambda}^{\circ}$ of the dual problem
in (\ref{eq:dualP}). Then the optimal primal solution of (\ref{eq:Pitert})
is given by $\boldsymbol{x}^{\circ}\left(\boldsymbol{\lambda}^{\circ}\right)$.

\subsection{Implementation Consideration}

At the beginning of each super-frame, the BS resets the SSCA-THP algorithm
with an initial THP variable $\boldsymbol{x}^{0}$. Then the THP variable
$\boldsymbol{x}$ is updated once every frame. Therefore, each frame
corresponds to an iteration in the SSCA-THP algorithm. Specifically,
let $\boldsymbol{x}^{l}=\left[\left(\boldsymbol{\phi}^{l}\right)^{T},\left(\boldsymbol{p}^{l}\right)^{T},\alpha^{l},\left(\boldsymbol{\beta}^{l}\right)^{T}\right]^{T}$
denote the THP variable used during the $l$-th frame. At time slot
$t$ in the $l$-th frame, the BS first acquires the effective channel
$\boldsymbol{H}\left(t\right)\boldsymbol{F}^{l}$, where $\boldsymbol{H}\left(t\right)$
is the channel state at time slot $t$, and $\boldsymbol{F}^{l}$
is the RF precoder corresponding to $\boldsymbol{\phi}^{l}$. Then
it calculates the baseband precoder $\boldsymbol{G}\left(t\right)$
from $\boldsymbol{H}\left(t\right)\boldsymbol{F}^{l}$ and $\boldsymbol{p}^{l},\alpha^{l}$
using (\ref{eq:Geq}). At the end of the $l$-th frame, the BS obtains
a channel sample $\boldsymbol{H}^{l}$ and updates the THP variable
$\boldsymbol{x}$ by solving a simple quadratic optimization problem,
where the updated THP variable $\boldsymbol{x}^{l+1}$ will be used
in the $\left(l+1\right)$-th frame. Then the same procedure is carried
out in the next frame.
\begin{rem}
The proposed SSCA-THP algorithm exploits some unique properties of
hybrid beamforming massive MIMO systems to improve the performance
and reduce the complexity. For example, the optimization variables
and constraints in Section \ref{subsec:Various-Implementation-Methods}
are specifically designed for hybrid beamforming with different RF
precoding structures. By imposing the sparse constraint in (\ref{eq:sparsecond}),
the property of limited RF chains in massive MIMO is also exploited
to improve the performance over the existing semidefinite relaxation
(SDR) method in \cite{Liu_TSP2016_CSImassive} for the codebook-based
RF precoder. Finally, the structure of the average data rate function
$\overline{r}_{k}\left(\boldsymbol{\theta},\boldsymbol{p},\alpha\right)$
w.r.t. the phase vector $\boldsymbol{\theta}$ is exploited to design
quadratic surrogate functions which enables low-complexity and fast-convergent
algorithm design.
\end{rem}

\section{Convergence Analysis\label{sec:Convergence-Analysis}}

In this section, we establish the local convergence of SSCA-THP to
a stationary point. There are several challenges in the convergence
proof for SSCA-THP. First, we need to show that at every limiting
point, all constraints are satisfied, which is non-trivial since SSCA-THP
may oscillate between the feasible update and objective update. Moreover,
the limiting point is obtained by averaging over all the previous
outputs from either feasible updates or objective updates, which makes
it difficult to show that the limiting point is a stationary point
of the original Problem (\ref{eq:mainP}). In this subsection, we
will overcome these challenges and establish the convergence of SSCA-THP.
To prove the convergence of SSCA-THP, we need to make the following
assumptions on the problem structure.

\begin{assumption}[Assumptions on Problem (\ref{eq:mainP})]\label{asm:convP}$\:$
\begin{enumerate}
\item $h_{i}\left(\overline{\boldsymbol{r}},\boldsymbol{x}\right),i=0,...,m$
are continuously differentiable functions of $\left(\overline{\boldsymbol{r}},\boldsymbol{x}\right)$.
\item For any $\boldsymbol{x}\in\mathcal{X}$, the functions $h_{i}\left(\overline{\boldsymbol{r}}\left(\boldsymbol{\phi},\boldsymbol{p},\alpha\right),\boldsymbol{x}\right),i=0,...,m$,
their derivative, and their second-order derivative w.r.t. $\overline{\boldsymbol{r}}$
and $\boldsymbol{x}$ are uniformly bounded.
\item $\left\Vert \boldsymbol{H}^{l}\right\Vert ,l=0,1,...$ are uniformly
bounded w.p.1.
\item Let $\boldsymbol{x}_{F}^{*}$ be any stationary point of the following
feasibility problem:
\begin{align}
\min_{\boldsymbol{x}\in\mathcal{X},\nu}\: & \nu\label{eq:FP}\\
s.t.\: & f_{i}\left(\boldsymbol{x}\right)\leq\nu,\:\forall i=1,....,m.\nonumber 
\end{align}
We assume that $f_{i}\left(\boldsymbol{x}_{F}^{*}\right)\leq0,i=1,...,m$. 
\end{enumerate}
\end{assumption}

The first assumption is standard and is satisfied for a large class
of problems. In practice, the channel sample is always bounded, and
thus the second assumption is satisfied. The third assumption ensures
that Problem (\ref{eq:mainP}) is feasible. If there is a stationary
point $\boldsymbol{x}_{F}^{*}$ which is not feasible, then Algorithm
1 may get stuck at this stationary point $\boldsymbol{x}_{F}^{*}$.
Therefore, the third assumption is necessary for the algorithm to
converge to a feasible point of the problem.

Besides Assumption \ref{asm:convP}, the sequence of parameters $\left\{ \rho^{t}\right\} ,\left\{ \gamma^{t}\right\} $
needs to satisfy the following conditions.

\begin{assumption}[Assumptions on $\left\{ \rho^{t}\right\} ,\left\{ \gamma^{t}\right\} $]\label{asm:convS}$\:$
\begin{enumerate}
\item $\rho^{l}\rightarrow0$, $\sum_{l}\rho^{l}=\infty$, $\sum_{l}\left(\rho^{l}\right)^{2}<\infty$,
$\lim_{l\rightarrow\infty}\rho^{l}l^{-1/2}<\infty$.
\item $\gamma^{l}\rightarrow0$, $\sum_{l}\gamma^{l}=\infty$, $\sum_{l}\left(\gamma^{l}\right)^{2}<\infty$,
\item $\lim_{l\rightarrow\infty}\gamma^{l}/\rho^{l}=0$.
\end{enumerate}
\end{assumption}

With Assumptions \ref{asm:convP} and \ref{asm:convS}, we can prove
two key lemmas that will eventually lead to the final convergence
result. The first lemma proves the convergence of surrogate functions.
\begin{lem}
[Convergence of surrogate functions]\label{lem:Convergence-surrogate}Suppose
Assumptions \ref{asm:convP} and \ref{asm:convS} are satisfied. Consider
a subsequence $\left\{ \boldsymbol{x}^{l_{j}}\right\} _{j=1}^{\infty}$
converging to a limiting point $\boldsymbol{x}^{*}$, and define functions
\begin{align*}
\hat{f}_{i}\left(\boldsymbol{x}\right) & \triangleq h_{i}\left(\overline{\boldsymbol{r}}\left(\boldsymbol{\phi}^{*},\boldsymbol{p}^{*},\alpha^{*}\right),\boldsymbol{x}^{*}\right)\\
 & +\nabla f_{i}\left(\boldsymbol{x}^{*}\right)\left(\boldsymbol{x}-\boldsymbol{x}^{*}\right)+\tau_{i}\left\Vert \boldsymbol{x}-\boldsymbol{x}^{*}\right\Vert ^{2},\forall i,
\end{align*}
which satisfy $\hat{f}_{i}\left(\boldsymbol{x}^{*}\right)=f_{i}\left(\boldsymbol{x}^{*}\right)$
and $\nabla\hat{f}_{i}\left(\boldsymbol{x}^{*}\right)=\nabla f_{i}\left(\boldsymbol{x}^{*}\right),\forall i$.
Then, almost surely, we have

\begin{align}
\lim_{j\rightarrow\infty}\bar{f}_{i}^{l_{j}}\left(\boldsymbol{x}\right) & =\hat{f}_{i}\left(\boldsymbol{x}\right),\:\forall\boldsymbol{x}\in\mathcal{X}.\label{eq:ghfhead}
\end{align}

\end{lem}
Please refer to Appendix \ref{subsec:Proof-of-Lemmaconvf} for the
proof. To state the convergence result, we need to introduce the concept
of Slater condition for the converged surrogate functions.

\textbf{Slater condition for the converged surrogate functions:} Given
a subsequence $\left\{ \boldsymbol{x}^{l_{j}}\right\} _{j=1}^{\infty}$
converging to a limiting point $\boldsymbol{x}^{*}$ and letting $\hat{f}_{i}\left(\boldsymbol{x}\right),\forall i$
be the converged surrogate functions as defined in Lemma \ref{lem:Convergence-surrogate},
we say that the Slater condition is satisfied at $\boldsymbol{x}^{*}$
if there exists $\boldsymbol{x}\in\textrm{int}\mathcal{X}$ such that
\[
\hat{f}_{i}\left(\boldsymbol{x}\right)<0,\:\forall i=1,...,m.
\]
A similar Slater condition is also assumed in \cite{Meisam_thesis14_BSUM}
to prove the convergence of a deterministic majorization-minimization
(MM) algorithm with non-convex constraints.

Before the introduction of the main convergence theorem, we give the
second key lemma. 
\begin{lem}
\label{lem:keylem}Let $\left\{ \boldsymbol{x}^{l}\right\} _{l=1}^{\infty}$
denote the sequence of iterates generated by Algorithm 1. We have
\begin{align*}
\limsup_{l\rightarrow\infty}\max_{i\in\left\{ 1,...,m\right\} }f_{i}\left(\boldsymbol{x}^{l}\right) & \leq0,\text{ w.p.1.}\\
\lim_{l\rightarrow\infty}\left\Vert \bar{\boldsymbol{x}}^{l}-\boldsymbol{x}^{l}\right\Vert  & =0,\text{ w.p.1.}
\end{align*}

\end{lem}
The lemma states that the algorithm will converge to the feasible
region, and the gap between $\bar{\boldsymbol{x}}^{l}$ and $\boldsymbol{x}^{l}$
converges to zero, almost surely. Please refer to Appendix \ref{subsec:Proof-of-keyLemma}
for the proof.
\begin{thm}
[Convergence of Algorithm 1]\label{thm:Convergence-of-Algorithm1}Suppose
Assumptions \ref{asm:convP} and \ref{asm:convS} are satisfied. For
any subsequence $\left\{ \boldsymbol{x}^{l_{j}}\right\} _{j=1}^{\infty}$
converging to a limit point $\boldsymbol{x}^{*}$, if the Slater condition
is satisfied at $\boldsymbol{x}^{*}$, then $\boldsymbol{x}^{*}$
is a stationary point of Problem (\ref{eq:mainP}) almost surely.
\end{thm}

Please refer to Appendix \ref{subsec:Proof-of-Theorem} for the proof.

\section{Applications and Numerical Validation\label{sec:Applications-and-Numerical}}

In this section, we shall apply the proposed SSCA-THP to solve the
first three example problems described in Section \ref{sec:System-Model}.
As in \cite{Park_TSP17_THP}, we adopt a geometry-based channel model
with a \textit{half-wavelength space} ULA for simulations. The channel
vector of user $k$ can be expressed as $\boldsymbol{h}_{k}=\sum_{i=1}^{N_{p}}\alpha_{k,i}\boldsymbol{\textrm{a}}\left(\varphi_{k,i}\right)$,
where $\boldsymbol{\textrm{a}}\left(\varphi\right)$ is the array
response vector, $\varphi_{k,i}$'s are Laplacian distributed with
an angle spread $\sigma_{\textrm{AS}}=10$, $\alpha_{k,i}\sim\mathcal{CN}\left(0,\sigma_{k,i}^{2}\right)$,
$\sigma_{k,i}^{2}$ are randomly generated from an exponential distribution
and normalized such that $\sum_{i=1}^{N_{p}}\sigma_{k,i}^{2}=g_{k}$,
and $g_{k}$ represents the path gain of user $k$. Unless otherwise
specified, we consider $M=64$ antennas, $S=16$ RF chains and $N_{p}=6$
channel paths. The path gains $g_{k}$'s are uniformly generated between
-10 dB and 10 dB. The DPS-based RF precoder is considered in Example
1 and 2, while both the DPS-based and codebook-based RF precoders
are considered in Example 3. We compare the performance of the SSCA-THP
with the following baseline algorithms.

Baseline 1 (SAA): This is the sample average approximation algorithm.
Specifically, after applying SAA, the problem becomes a deterministic
non-convex optimization problem, which is then solved using the deterministic
successive convex approximation method \cite{Meisam_thesis14_BSUM}. 

Baseline 2 (SLNR-max): This is the SLNR maximization algorithm in
\cite{Park_TSP17_THP}.

Baseline 3 (JSDM): This is the the joint spatial division and multiplexing
scheme in \cite{Caire_TIT13_JSDM}.

Baseline 4 (CB): This is the CB algorithm in \cite{Liu_TSP14_RFprecoding},
which combines the deterministic approximation and bisection methods. 

Both SSCA-THP and SAA can be used to solve a general THP optimization
problem, while the SLNR-max/JSDM is more suitable for the sum throughput
maximization in Example 1, and the CB algorithm can be used to solve
the power minimization problem in Example 3. Both the SLNR-max and
CB algorithms only work for the fully-connected RF precoding structure.
For fair comparison, we focus on the fully-connected structure in
the simulations. All baseline algorithms belong to the offline method,
which requires a channel sample collection phase to construct the
SAA functions (baseline 1) or estimate the channel covariance matrices
(baseline 2 and 3). We assume that one super-frame has $L=1000$ frames
and the first 200 frames serve as the channel sample collection phase
for the baseline algorithms. For fair comparison, the proposed SSCA-THP
is also terminated after 200 frames (iterations). The performance
is obtained by averaging over the last 800 frames of the super-frame.
Note that if we considered the overall performance averaged over the
entire super-frame, the proposed SSCA-THP would achieve an even larger
performance gain over the baseline algorithms, which perform poorly
during the channel sample collection phase.

\subsection{Convergence of the Proposed SSCA-THP}

\begin{figure}
\begin{centering}
\includegraphics[clip,width=80mm]{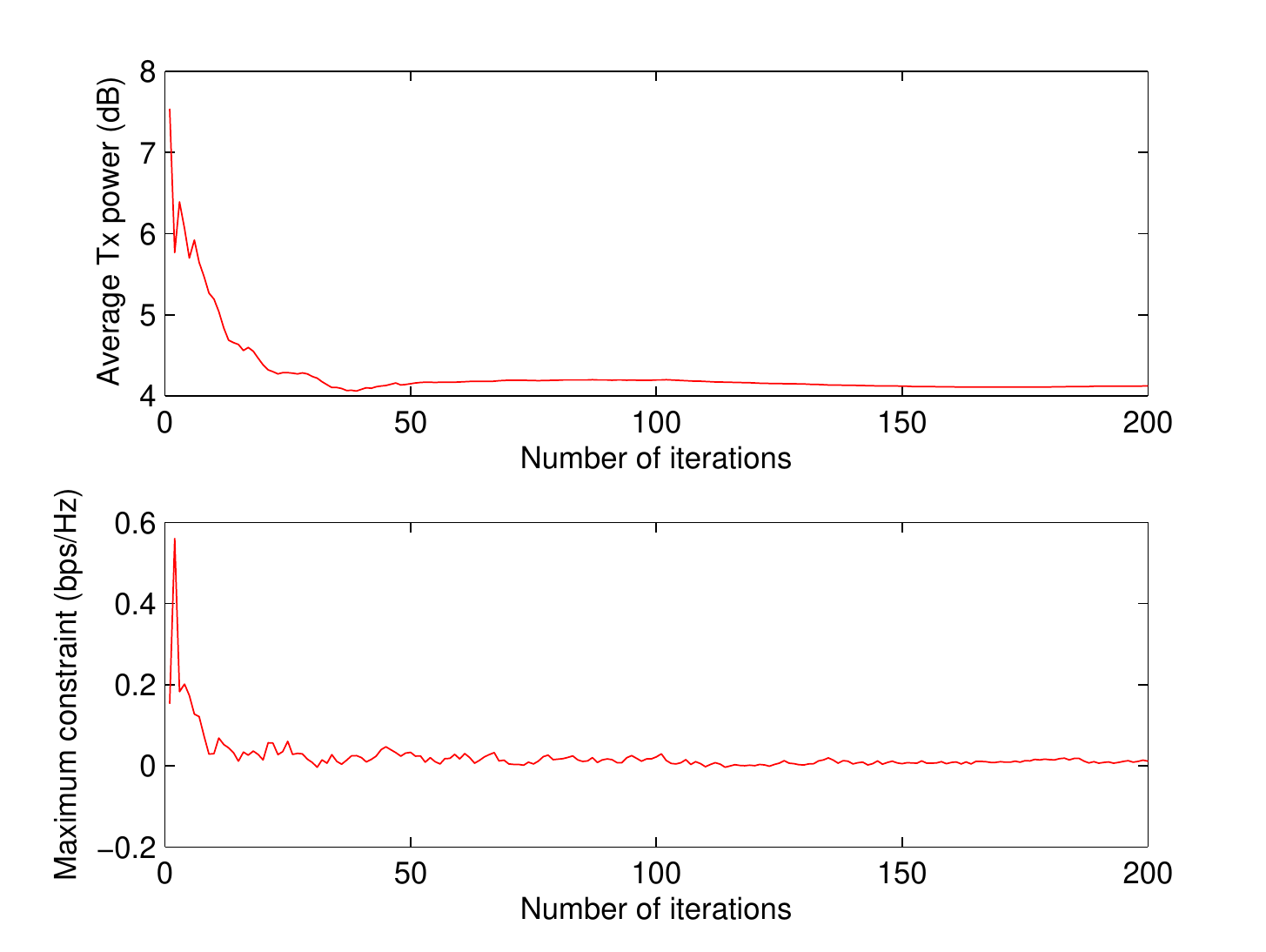}
\par\end{centering}
\caption{\label{fig:convg}{\small{}Convergence of the SSCA-THP}}
\end{figure}

We use Example 3 to illustrate the fast convergence of SSCA-THP. Specifically,
there are $K=8$ users and the target average rate for all users is
set to be the same as $\gamma_{k}=2$ bps/Hz. Consider the DFT-based
RF precoder. In Fig. \ref{fig:convg}, we plot the objective function
(average transmit power) and the maximum constraint function (target
average rate minus the minimum achieved average rate of users) versus
the iteration number, respectively. It can be seen that SSCA-THP quickly
converges to a point with all target average rates satisfied with
high accuracy.

\subsection{Sum Throughput Maximization}

\begin{figure}
\begin{centering}
\includegraphics[clip,width=80mm]{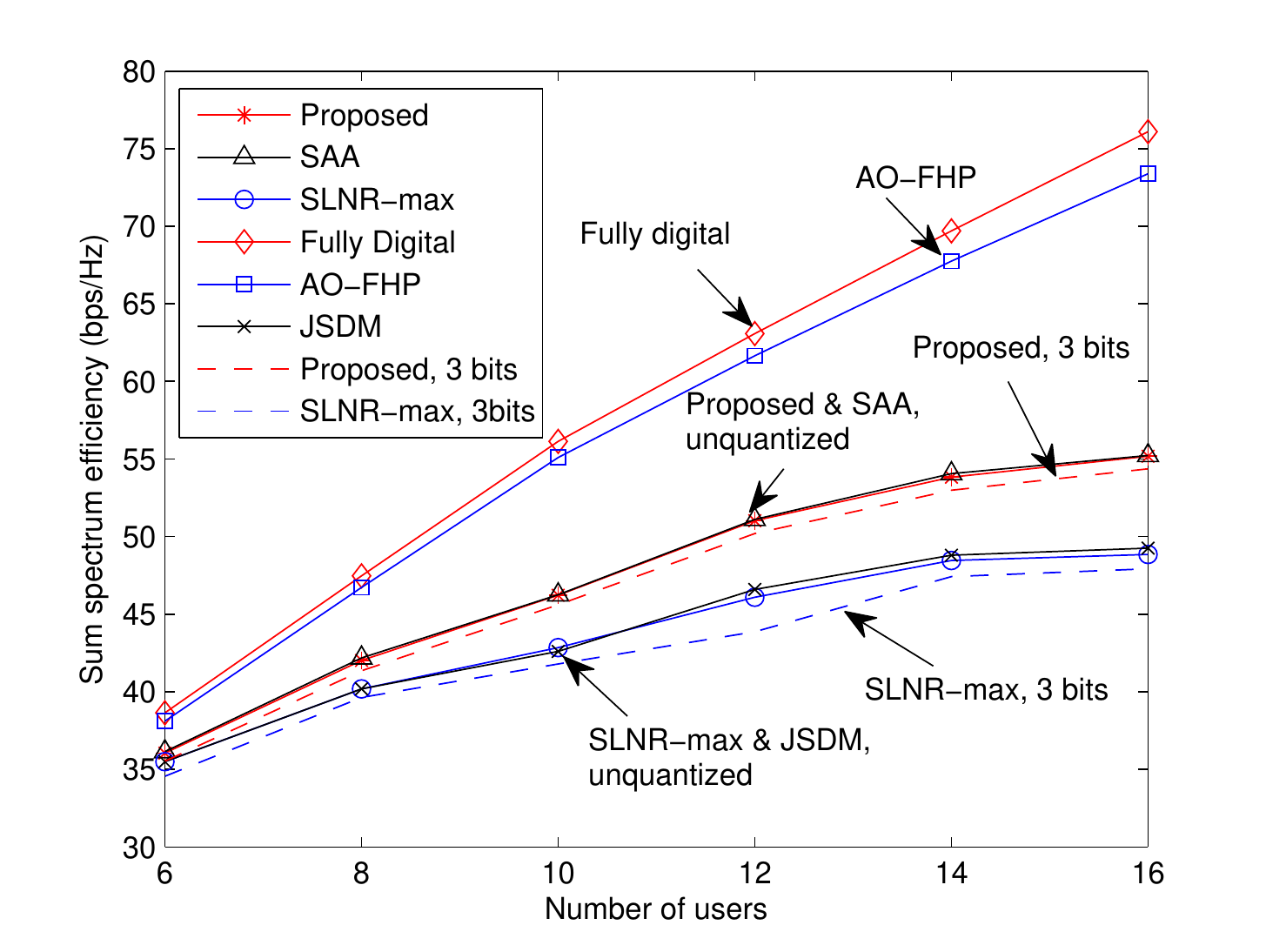}
\par\end{centering}
\caption{\label{fig:exm1}{\small{}Sum throughput (bps/Hz) versus the number
of users $K$}}
\end{figure}

\begin{figure}
\begin{centering}
\includegraphics[clip,width=80mm]{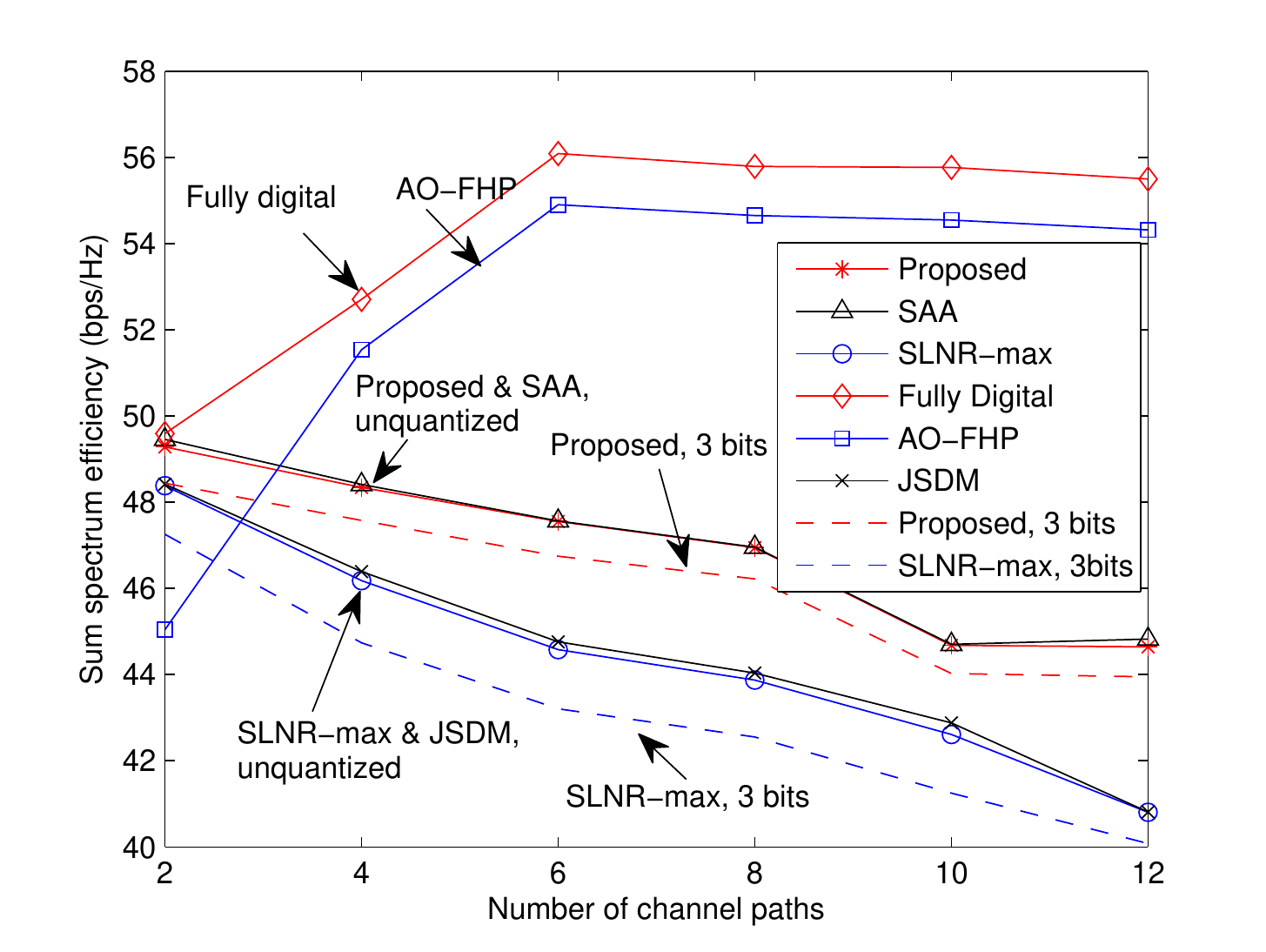}
\par\end{centering}
\caption{\label{fig:exm1-1}{\small{}Sum throughput versus the number of channel
paths $N_{p}$}}
\end{figure}

In Figs. \ref{fig:exm1} and \ref{fig:exm1-1}, we plot the sum throughput
versus the number of users $K$ and the number of channel paths $N_{p}$,
respectively. The transmit power is set to be $P=10$ dB. For comparison,
we also plot the sum throughput of the fully digital RZF beamforming
\cite{Peel_TOC05_RCI} and the FHP algorithm based on alternating
optimization (AO-FHP) in \cite{Yu_JSTSP2016_HP}. The proposed SSCA-THP
achieves better performance than the existing THP algorithms (SLNR-max
and JSDM). Moreover, as the number of users increases, the performance
gap between SSCA-THP and SLNR-max/JSDM increases. Although SAA achieves
similar performance to SSCA-THP after the channel sample collection
phase, it has poor performance during the channel sample collection
phase. Moreover, the per iteration complexity of SAA is much higher
than that of SSCA-THP (CPU time: 10.00 s versus 0.06 s). The performance
gap between the THP with statistical RF precoder and the FHP with
real-time RF precoder (or fully digital RZF) is smaller when the number
of users/channel paths is smaller. This is consistent with the analysis
in \cite{Liu_TSP2016_CSImassive}. Finally, it can be seen that with
only three-bit quantization, the performance is already very close
to the case without quantization.

\subsection{Proportional Fairness}

\begin{figure}
\begin{centering}
\includegraphics[clip,width=80mm]{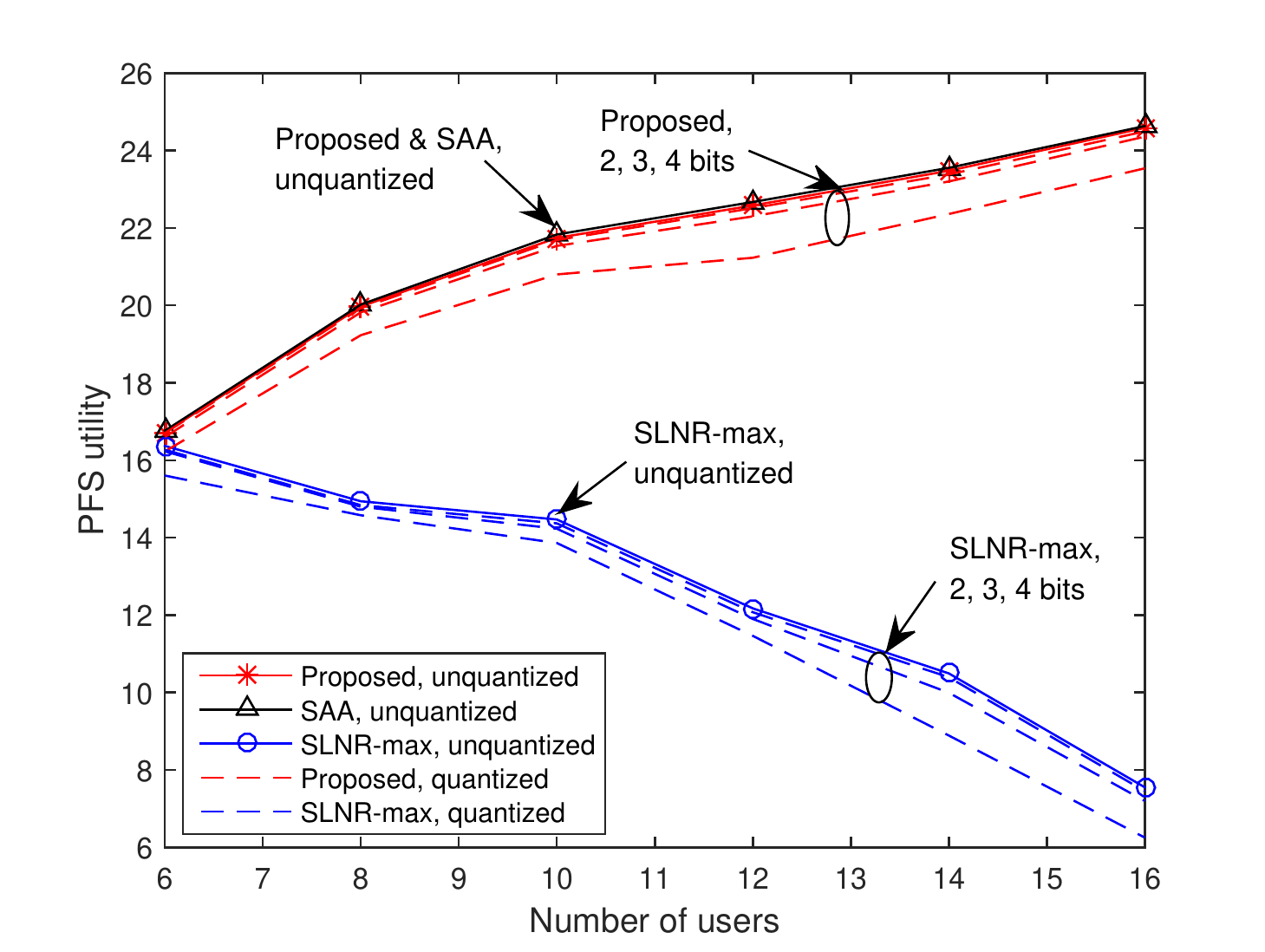}
\par\end{centering}
\caption{\label{fig:exm2a}{\small{}PFS utility versus the number of users
$K$}}
\end{figure}

In Fig. \ref{fig:exm2a}, we plot the PFS utility versus the number
of users $K$. The transmit power is set to be $P=10$ dB. Similar
results to those in Fig. \ref{fig:exm1} can be observed. Moreover,
when considering the PFS utility, the performance gap between SSCA-THP
and SLNR-max is much larger since the fairness issue is not considered
in the SLNR-max algorithm. Note that the PFS utility of SLNR-max decreases
with the number of users. This is because, without considering the
fairness, the minimum throughput of the users becomes much smaller
as the number of users increases.

\subsection{Power Minimization}

\begin{figure}
\begin{centering}
\includegraphics[clip,width=80mm]{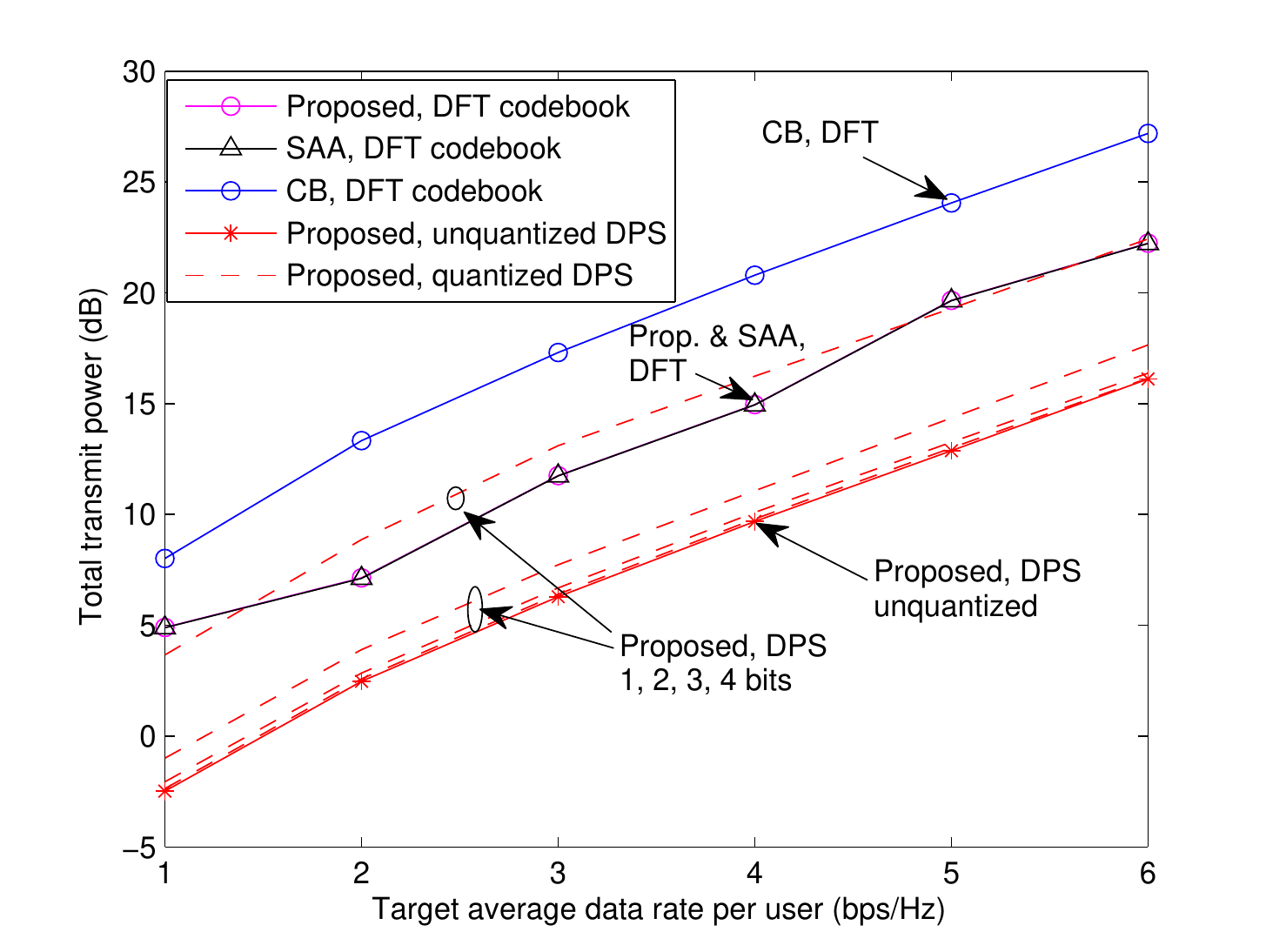}
\par\end{centering}
\caption{\label{fig:exm3}{\small{}Average transmit power versus the target
average rate}}
\end{figure}

In Fig. \ref{fig:exm3}, we plot the average transmit power versus
the target average rate requirement $\gamma_{k}=\gamma,\forall k$
for a system with $K=8$ users. For any given target average rate
$\gamma$, the proposed SSCA-THP with the DPS-based RF precoder achieves
the lowest transmit power. The performance of the DPS-based RF precoder
degrades as the number of quantization bits $B$ for each phase decreases.
When $B=1$, the performance of the DPS-based RF precoder is similar
to that of the DFT-based RF precoder optimized using the SSCA-THP
or SAA algorithms, which is still much better than the DFT-based RF
precoder optimized using the CB algorithm.

\subsection{Complexity Comparison}

The complexity of SSCA-THP is dominated by the calculation of the
Jacobian matrix $\mathbf{J}_{r}\left(\boldsymbol{x}^{l};\boldsymbol{H}^{l}\right)$,
which has complexity order $O\left(MKS\right)$, and the quadratic
optimization subproblems in (\ref{eq:Pitert}) and (\ref{eq:Pitert-1}),
which has complexity order $O\left(Mm^{2}\right)$, as explained below.
For given Lagrange multipliers, the complexity order of calculating
the closed-form primal solution in (\ref{eq:primopt}) is $O\left(M\right)$.
Using the ellipsoid method, the number of iterations required to achieve
a given convergence accuracy $\epsilon$ for the dual problem (\ref{eq:dualP})
is $O\left(m^{2}\log\left(1/\epsilon\right)\right)$ \cite{Boyd_04Book_Convex_optimization}.
Hence, the per-iteration complexity of SSCA-THP is $O\left(Mm^{2}\log\frac{1}{\epsilon}+MKS\right)$.
In Table \ref{tab:Cputime}, we compare the complexity order of SSCA-THP
with SLNR-max in \cite{Park_TSP17_THP} and JSDM in \cite{Caire_TIT13_JSDM}
for the sum throughput maximization problem (i.e., $m=1$), where
$L_{I}$ is the total number of iterations for SSCA-THP and Algorithm
1 in \cite{Park_TSP17_THP}, respectively. The complexity order of
both SLNR-max and JSDM increases with the number of BS antennas $M$
according to $M^{3}$ because they involve singular value decomposition
(SVD) for the $M\times M$ channel covariance matrix. On the other
hand, the complexity order of SSCA-THP only increases linearly with
$M$, thanks to the closed-form solution in (\ref{eq:primopt}) for
fixed Lagrange multipliers.

\begin{table}
\begin{centering}
{\footnotesize{}}%
\begin{tabular}{|l|l|l|}
\hline 
SSCA-THP & SLNR-max & JSDM\tabularnewline
\hline 
{\small{}$O\left(L_{I}\left(M\log\frac{1}{\epsilon}+MKS\right)\right)$} & {\small{}$O\left(L_{I}M^{2}S+M^{3}\right)$} & {\small{}$O\left(M^{3}\right)$}\tabularnewline
\hline 
\end{tabular}
\par\end{centering}{\footnotesize \par}
{\small{}\caption{\label{tab:Cputime}{\small{}Comparison of the complexity order for
different algorithms.}}
}{\small \par}
\end{table}

\section{Conclusion\label{sec:Conlusion}}

In this paper, we first propose a general optimization formulation
(\ref{eq:mainP}) for the design of THP in massive MIMO, which is
applicable to different RF precoding structures/implementations and
a wide range of application scenarios. Then we propose an online algorithmic
framework called SSCA-THP to solve this general THP optimization problem.
Specifically, at each iteration, quadratic surrogate functions are
constructed for both objective and constraint functions based on a
new channel sample. Then the next iterate is updated by solving the
resulting quadratic optimization problem. We prove the convergence
of SSCA-THP to stationary points. To the best of our knowledge, SSCA-THP
is the first online and provably convergent algorithm to handle the
general non-convex stochastic constraints considered in (\ref{eq:mainP}).
Finally, we apply SSCA-THP to solve three important THP optimization
problems and verify its advantages.

\appendix

\subsection{Jacobian Matrix of Instantaneous Rate\label{subsec:Jacobian-Matrix-of} }

\subsubsection{Jacobian Matrix for the Fully-connected DPS-based RF Precoder}

In this case, we have $\boldsymbol{\phi}=\boldsymbol{\theta}=\mathbb{R}^{MS}$
and $\boldsymbol{x}=\left[\boldsymbol{\theta}^{T},\boldsymbol{p}^{T},\alpha,\boldsymbol{\beta}^{T}\right]^{T}$.
We first define some useful notations:
\begin{align*}
\boldsymbol{A}_{k,i} & =2\boldsymbol{H}_{F}^{H}\mathfrak{S}\left[\boldsymbol{H}\boldsymbol{F}\boldsymbol{F}^{H}\boldsymbol{h}_{k}\boldsymbol{h}_{k}^{H}\overline{\boldsymbol{G}}\boldsymbol{P}_{i}\boldsymbol{\Lambda}\right]\boldsymbol{H}_{F}\boldsymbol{F}\\
 & -2\mathfrak{S}\left[\boldsymbol{h}_{k}\boldsymbol{h}_{k}^{H}\overline{\boldsymbol{G}}\boldsymbol{P}_{i}\boldsymbol{\Lambda}\boldsymbol{H}_{F}\right]\boldsymbol{F},\\
\boldsymbol{E}_{i} & =2\boldsymbol{H}_{F}^{H}\mathfrak{S}\left[\boldsymbol{H}\boldsymbol{F}\boldsymbol{F}^{H}\overline{\boldsymbol{G}}\boldsymbol{I}_{i}\right]\boldsymbol{H}_{F}\boldsymbol{F}-2\mathfrak{S}\left[\overline{\boldsymbol{G}}\boldsymbol{I}_{i}\boldsymbol{H}_{F}\right]\boldsymbol{F},\\
e_{k,i} & =\left[\boldsymbol{H}_{F}\boldsymbol{F}\boldsymbol{F}^{H}\boldsymbol{h}_{k}\boldsymbol{h}_{k}^{H}\boldsymbol{F}\boldsymbol{F}^{H}\boldsymbol{H}_{F}^{H}\boldsymbol{P}_{i}\boldsymbol{\Lambda}^{2}\right]_{i,i},
\end{align*}
where $\boldsymbol{H}_{F}=\boldsymbol{B}\boldsymbol{H}$ with $\boldsymbol{B}=\left(\boldsymbol{H}\boldsymbol{F}\boldsymbol{F}^{H}\boldsymbol{H}^{H}+\alpha\boldsymbol{I}\right)^{-1}$,
and $\boldsymbol{P}_{i}$ ($\boldsymbol{I}_{i}$) denotes a $K\times K$
matrix with $\left[\boldsymbol{P}_{i}\right]_{i,i}=p_{i}$ ($\left[\boldsymbol{I}_{i}\right]_{i,i}=1$)
and all other elements being zero. Then using the matrix calculus,
it can be shown that the gradients of $r_{k}\left(\boldsymbol{\theta},\boldsymbol{p},\alpha;\boldsymbol{H}\right)$
w.r.t. $\boldsymbol{\theta}$, $\boldsymbol{p}$ and $\alpha$ are
respectively given by
\begin{align}
\nabla_{\boldsymbol{\theta}}r_{k}\left(\boldsymbol{\theta},\boldsymbol{p},\alpha;\boldsymbol{H}\right) & =\frac{\sum_{i}\boldsymbol{a}_{k,i}^{\theta}}{\varGamma_{k}}-\frac{\sum_{i\neq k}\boldsymbol{a}_{k,i}^{\theta}}{\varGamma_{-k}}\label{Grtheta}\\
\nabla_{\boldsymbol{p}}r_{k}\left(\boldsymbol{\theta},\boldsymbol{p},\alpha;\boldsymbol{H}\right) & =\frac{\sum_{i}\boldsymbol{a}_{k,i}^{p}}{\varGamma_{k}}-\frac{\sum_{i\neq k}\boldsymbol{a}_{k,i}^{p}}{\varGamma_{-k}},\label{Grp}\\
\nabla_{\alpha}r_{k}\left(\boldsymbol{\theta},\boldsymbol{p},\alpha;\boldsymbol{H}\right) & =\frac{\sum_{i}a_{k,i}^{\alpha}}{\varGamma_{k}}-\frac{\sum_{i\neq k}a_{k,i}^{\alpha}}{\varGamma_{-k}},\label{eq:Gralpha}
\end{align}
where $\varGamma_{k}=\sum_{i}p_{i}\left|\boldsymbol{h}_{k}^{H}\boldsymbol{F}\boldsymbol{g}_{i}\right|^{2}+1$,
$\varGamma_{-k}=\sum_{i\neq k}p_{i}\left|\boldsymbol{h}_{k}^{H}\boldsymbol{F}\boldsymbol{g}_{i}\right|^{2}+1$,
{\small{}
\begin{align*}
\boldsymbol{a}_{k,i}^{\theta} & =\textrm{Vec}\left(\mathfrak{R}\left[\sqrt{-1}\boldsymbol{F}^{*}\circ\boldsymbol{A}_{k,i}\right]\right)-e_{k,i}\textrm{Vec}\left(\mathfrak{R}\left[\sqrt{-1}\boldsymbol{F}^{*}\circ\boldsymbol{E}_{i}\right]\right),\\
\boldsymbol{a}_{k,i}^{p} & =\textrm{Diag}\left(\overline{\boldsymbol{G}}^{H}\boldsymbol{h}_{k}\boldsymbol{h}_{k}^{H}\overline{\boldsymbol{G}}\boldsymbol{\Lambda}_{i}\right),\\
a_{k,i}^{\alpha} & =e_{k,i}2\mathfrak{R}\left[Tr\left(\overline{\boldsymbol{G}}\boldsymbol{I}_{i}\boldsymbol{B}\overline{\boldsymbol{G}}^{H}\right)\right]-2\mathfrak{R}\left[\boldsymbol{h}_{k}^{H}\overline{\boldsymbol{G}}\boldsymbol{P}_{i}\boldsymbol{\Lambda}_{i}\boldsymbol{B}\overline{\boldsymbol{G}}^{H}\boldsymbol{h}_{k}\right].
\end{align*}
}Therefore, for given channel state $\boldsymbol{H}$, the Jacobian
matrix of the instantaneous rate vector $\boldsymbol{r}\left(\boldsymbol{\theta},\boldsymbol{p},\alpha;\boldsymbol{H}\right)$
w.r.t. $\boldsymbol{x}$ is
\begin{equation}
\mathbf{J}_{r}\left(\boldsymbol{x};\boldsymbol{H}\right)=\left[\begin{array}{cccc}
\nabla_{\boldsymbol{\theta}}r_{1} & \nabla_{\boldsymbol{\theta}}r_{2} & \cdots & \nabla_{\boldsymbol{\theta}}r_{K}\\
\nabla_{\boldsymbol{p}}r_{1} & \nabla_{\boldsymbol{p}}r_{2} & \cdots & \nabla_{\boldsymbol{p}}r_{K}\\
\nabla_{\alpha}r_{1} & \nabla_{\alpha}r_{2} & \cdots & \nabla_{\alpha}r_{K}\\
\boldsymbol{0} & \boldsymbol{0} & \boldsymbol{0} & \boldsymbol{0}
\end{array}\right],\label{eq:Jrx}
\end{equation}
where the bottom submatrix is zero because $\nabla_{\boldsymbol{\beta}}r_{k}=\boldsymbol{0},\forall k$.
Note that we have omitted $\left(\boldsymbol{\theta},\boldsymbol{p},\alpha;\boldsymbol{H}\right)$
in the gradient expressions for simplicity of notation.

\subsubsection{Jacobian Matrix for the Fully-connected Codebook-based RF Precoder}

In this case, we have $\boldsymbol{\phi}=\boldsymbol{d}=\left[0,1\right]^{N}$
and $\boldsymbol{x}=\left[\boldsymbol{d}^{T},\boldsymbol{p}^{T},\alpha,\boldsymbol{\beta}^{T}\right]^{T}$.
Using the matrix calculus, it can be shown that the gradients of $r_{k}\left(\boldsymbol{d},\boldsymbol{p},\alpha;\boldsymbol{H}\right)$
w.r.t. $\boldsymbol{p}$ and $\alpha$ are given by (\ref{Grp}) and
(\ref{eq:Gralpha}), respectively; and the gradient of $r_{k}\left(\boldsymbol{d},\boldsymbol{p},\alpha;\boldsymbol{H}\right)$
w.r.t. $\boldsymbol{d}$ is
\[
\nabla_{\boldsymbol{d}}r_{k}\left(\boldsymbol{\theta},\boldsymbol{p},\alpha;\boldsymbol{H}\right)=\frac{\sum_{i}\boldsymbol{a}_{k,i}^{d}}{\varGamma}-\frac{\sum_{i\neq k}\boldsymbol{a}_{k,i}^{d}}{\varGamma_{k}},
\]
where $\boldsymbol{a}_{k,i}^{d}=-\frac{1}{2}\textrm{Diag}\left(\boldsymbol{C}^{H}\boldsymbol{A}_{k,i}\boldsymbol{C}\right)+\frac{e_{k,i}}{2}\textrm{Diag}\left(\boldsymbol{C}^{H}\boldsymbol{E}_{i}\boldsymbol{C}\right)$.
Finally, for given channel state $\boldsymbol{H}$, the Jacobian matrix
of the instantaneous rate vector $\boldsymbol{r}\left(\boldsymbol{d},\boldsymbol{p},\alpha;\boldsymbol{H}\right)$
w.r.t. $\boldsymbol{x}$ is given by (\ref{eq:Jrx}) with $\nabla_{\boldsymbol{\theta}}r_{k},\forall k$
replaced by $\nabla_{\boldsymbol{d}}r_{k},\forall k$.

The Jacobian matrix of the instantaneous rate vector for the partially-connected
structure can be obtained similarly. The details are omitted for conciseness.

\subsection{Proof of Lemma \ref{lem:Convergence-surrogate}\label{subsec:Proof-of-Lemmaconvf}}

The proof relies on the following lemma.
\begin{lem}
\label{lem:fconv}Under Assumption \ref{asm:convS}, we have
\begin{align}
\lim_{l\rightarrow\infty}\left|\bar{f}_{i}^{l}\left(\boldsymbol{x}^{l}\right)-f_{i}\left(\boldsymbol{x}^{l}\right)\right| & =0,\label{eq:fconv1}\\
\lim_{l\rightarrow\infty}\left\Vert \nabla\bar{f}_{i}^{l}\left(\boldsymbol{x}^{l}\right)-\nabla f_{i}\left(\boldsymbol{x}^{l}\right)\right\Vert  & =0,\label{eq:fconv2}\\
\lim_{l_{1},l_{2}\rightarrow\infty}\bar{f}_{i}^{l_{1}}\left(\boldsymbol{x}^{l_{1}}\right)-\bar{f}_{i}^{l_{2}}\left(\boldsymbol{x}^{l_{2}}\right) & \leq C\left\Vert \boldsymbol{x}^{l_{1}}-\boldsymbol{x}^{l_{2}}\right\Vert .\label{eq:fconv3}
\end{align}
for $i=0,...,m$\textup{ w.p.1.}, where $C>0$ is some constant.
\end{lem}

\begin{IEEEproof}
It follows from the law of large numbers and the central limit theorem
that 
\begin{align}
\hat{\boldsymbol{r}}^{l}\overset{a.s.}{\rightarrow}\overline{\boldsymbol{r}}^{l},\: & \mathbb{E}\left\Vert \hat{\boldsymbol{r}}^{l}-\overline{\boldsymbol{r}}^{l}\right\Vert =O\left(\frac{1}{\sqrt{l}}\right),\label{eq:CLTK}
\end{align}
where $\overline{\boldsymbol{r}}^{l}=\overline{\boldsymbol{r}}\left(\boldsymbol{\phi}^{l},\boldsymbol{p}^{l},\alpha^{l}\right)$.
Then (\ref{eq:fconv1}) follows from (\ref{eq:CLTK}).

On the other hand, (\ref{eq:fconv2}) is a consequence of \cite{Ruszczyski_MP80_SPthem},
Lemma 1. It is easy to verify that the technical conditions (a), (b),
(d) and (e) therein are satisfied. In the following, we prove that
condition (c) in \cite{Ruszczyski_MP80_SPthem}, Lemma 1 is also satisfied.
Let $\overline{\nabla}_{\xi}^{l}h_{i}=\nabla_{\xi}h_{i}\left(\overline{\boldsymbol{r}}^{l},\boldsymbol{x}^{l}\right)$
and $\hat{\nabla}_{\xi}^{l}h_{i}=\nabla_{\xi}h_{i}\left(\hat{\boldsymbol{r}}^{l},\boldsymbol{x}^{l}\right)$
for $\xi\in\left\{ \overline{\boldsymbol{r}},\boldsymbol{x}\right\} $.
Let $\mathbf{J}_{\overline{r}}\left(\boldsymbol{x}^{l}\right)=\mathbb{E}\left[\mathbf{J}_{r}\left(\boldsymbol{x}^{l},\boldsymbol{H}^{l}\right)\right]$
denote the Jacobian matrix of the average rate vector $\overline{\boldsymbol{r}}\left(\boldsymbol{\phi}^{l},\boldsymbol{p}^{l},\alpha^{l}\right)$
at point $\boldsymbol{x}^{l}$. Then we have 
\begin{equation}
\nabla f_{i}\left(\boldsymbol{x}^{l}\right)=\mathbf{J}_{\overline{r}}\left(\boldsymbol{x}^{l}\right)\overline{\nabla}_{\overline{\boldsymbol{r}}}^{l}h_{i}+\overline{\nabla}_{\boldsymbol{x}}^{l}h_{i}.\label{eq:Gfxl}
\end{equation}
It follows from (\ref{eq:Gfxl}) and (\ref{eq:CLTK}) that

\begin{align}
\left\Vert \mathbb{E}\left[\hat{\mathbf{u}}_{i}^{l}\right]-\nabla f_{i}\left(\boldsymbol{x}^{l}\right)\right\Vert  & \leq\mathbb{E}\left\Vert \mathbf{J}_{r}\left(\boldsymbol{x}^{l},\boldsymbol{H}^{l}\right)\left(\hat{\nabla}_{\overline{\boldsymbol{r}}}^{l}h_{i}-\overline{\nabla}_{\overline{\boldsymbol{r}}}^{l}h_{i}\right)\right\Vert \nonumber \\
+ & \mathbb{E}\left\Vert \hat{\nabla}_{\boldsymbol{x}}^{l}h_{i}-\overline{\nabla}_{\boldsymbol{x}}^{l}h_{i}\right\Vert \nonumber \\
\overset{\textrm{a}}{=} & O\left(\left\Vert \hat{\boldsymbol{r}}^{l}-\overline{\boldsymbol{r}}^{l}\right\Vert \right)=O\left(\frac{1}{\sqrt{l}}\right),\label{eq:Euhed}
\end{align}
where (\ref{eq:Euhed}-a) holds because $\nabla h_{i}$ are Lipschitz
continuous and $\mathbf{J}_{r}\left(\boldsymbol{x}^{l},\boldsymbol{H}^{l}\right)$
are bounded w.p.1. From (\ref{eq:Euhed}) and $\sum_{l=1}^{\infty}\rho^{l}l^{-0.5}<\infty$,
we have $\sum_{l=1}^{\infty}\rho^{l}\left\Vert \mathbb{E}\left[\hat{\mathbf{u}}_{i}^{l}\right]-\nabla f_{i}\left(\boldsymbol{x}^{l}\right)\right\Vert <\infty$,
which implies that the technical condition (c) in \cite{Ruszczyski_MP80_SPthem},
Lemma 1 is satisfied. 

Finally, (\ref{eq:fconv3}) follows from the Lipschitz continuity
of $h_{i}$. This completes the proof.
\end{IEEEproof}

From Lemma \ref{lem:fconv} and (\ref{eq:CLTK}), the families of
functions $\left\{ \bar{f}_{i}^{l_{j}}\left(\boldsymbol{x}\right)\right\} $
converge to $\left\{ \hat{f}_{i}\left(\boldsymbol{x}\right)\right\} $
almost surely.

\subsection{Proof of Lemma \ref{lem:keylem}\label{subsec:Proof-of-keyLemma}}

1. We first prove $\limsup_{l\rightarrow\infty}f\left(\boldsymbol{x}^{l}\right)\leq0$
 w.p.1., where $f\left(\boldsymbol{x}\right)=\max_{i\in\left\{ 1,...,m\right\} }f_{i}\left(\boldsymbol{x}\right)$. 

Let $\mathcal{T}_{\epsilon}=\left\{ l:\:f\left(\boldsymbol{x}^{l}\right)\geq\epsilon\right\} $
for any $\epsilon>0$. We show that $\mathcal{T}_{\epsilon}$ is a
finite set by contradiction. 

Suppose $\mathcal{T}_{\epsilon}$ is infinite. We first show that
$\liminf_{l\in\mathcal{T}_{\epsilon},l\rightarrow\infty}\left\Vert \bar{\boldsymbol{x}}^{l}-\boldsymbol{x}^{l}\right\Vert >0$
by contradiction. Suppose $\liminf_{l\in\mathcal{T}_{\epsilon},l\rightarrow\infty}\left\Vert \bar{\boldsymbol{x}}^{l}-\boldsymbol{x}^{l}\right\Vert =0$.
Then there exists a subsequence $l^{j}\in\mathcal{T}_{\epsilon}$
such that $\lim_{j\rightarrow\infty}\left\Vert \bar{\boldsymbol{x}}^{l_{j}}-\boldsymbol{x}^{l_{j}}\right\Vert =0$.
Let $\boldsymbol{x}^{\circ}$ denote a limiting point of the subsequence
$\left\{ \boldsymbol{x}^{l_{j}}\right\} $, and let $\hat{f}_{i}\left(\boldsymbol{x}\right),\forall i$
be the converged surrogate functions as defined in Lemma \ref{lem:Convergence-surrogate}.
According to the update rule of Algorithm 1, there are two cases.

Case 1: $\boldsymbol{x}^{\circ}$ is the optimal solution of the following
convex optimization problem: 

\begin{align}
\underset{\boldsymbol{x}}{\text{min}}\: & \hat{f}_{0}\left(\boldsymbol{x}\right)\label{eq:Pcase1}\\
s.t.\: & \hat{f}_{i}\left(\boldsymbol{x}\right)\leq0,i=1,....,m.\nonumber 
\end{align}
In this case, we have $f\left(\boldsymbol{x}^{\circ}\right)=\max_{i\in\left\{ 1,...,m\right\} }\hat{f}_{i}\left(\boldsymbol{x}^{\circ}\right)\leq0$,
which contradicts the definition of $\mathcal{T}_{\epsilon}$.

Case 2: $\boldsymbol{x}^{\circ}$ is the optimal solution of the following
convex optimization problem: 

\begin{align}
\underset{\boldsymbol{x},\nu}{\text{min}}\: & \nu\label{eq:Piterthead-1}\\
s.t.\: & \hat{f}_{i}\left(\boldsymbol{x}\right)\leq\nu,i=1,....,m.\nonumber 
\end{align}
Since the Slater condition is satisfied (by choosing a sufficiently
large $\nu$, we can always find a point $\boldsymbol{x}\in\mathcal{X}$
such that $\hat{f}_{i}\left(\boldsymbol{x}\right)<\nu,i=1,....,m$),
the KKT condition of the problem (\ref{eq:Piterthead-1}) implies
that there exist $\lambda_{1},...,\lambda_{m}$ such that
\begin{align}
\sum_{i}\lambda_{i}\nabla\hat{f}_{i}\left(\boldsymbol{x}^{\circ}\right) & =\boldsymbol{0},\nonumber \\
1-\sum_{i}\lambda_{i} & =0,\nonumber \\
\hat{f}_{i}\left(\boldsymbol{x}^{\circ}\right) & \leq\nu,\:\forall i=1,...,m,\nonumber \\
\lambda_{i}\left(\hat{f}_{i}\left(\boldsymbol{x}^{\circ}\right)-\nu\right) & =0,\:\forall i=1,...,m.\label{KKTPhead1}
\end{align}
It follows from Lemma \ref{lem:Convergence-surrogate} and (\ref{KKTPhead1})
that $\boldsymbol{x}^{\circ}$ also satisfies the KKT condition of
Problem (\ref{eq:FP}). By Assumption \ref{asm:convP}, we have $f_{i}\left(\boldsymbol{x}^{\circ}\right)\leq0,i=1,...,m$,
which again contradicts the definition of $\mathcal{T}_{\epsilon}$.

Therefore, $\liminf_{l\in\mathcal{T}_{\epsilon},l\rightarrow\infty}\left\Vert \bar{\boldsymbol{x}}^{l}-\boldsymbol{x}^{l}\right\Vert >0$;
i.e., there exists a sufficiently large $l_{\epsilon}$ such that
\begin{equation}
\left\Vert \bar{\boldsymbol{x}}^{l}-\boldsymbol{x}^{l}\right\Vert \geq\epsilon^{'},\forall l\in\mathcal{T}_{\epsilon}^{'},\label{eq:gapxt}
\end{equation}
where $\epsilon^{'}>0$ is some constant and $\mathcal{T}_{\epsilon}^{'}=\mathcal{T}_{\epsilon}\cap\left\{ l\geq l_{\epsilon}\right\} $. 

Define function $\bar{f}^{l}\left(\boldsymbol{x}\right)=\max_{i\in\left\{ 1,...,m\right\} }\bar{f}_{i}^{l}\left(\boldsymbol{x}\right)$.
From the definition of $\bar{f}_{i}^{l}\left(\boldsymbol{x}\right)$
in (\ref{eq:upsurrgate}), $\bar{f}_{i}^{l}\left(\boldsymbol{x}\right)$
is strongly convex, and thus
\begin{equation}
\nabla^{T}\bar{f}_{i}^{l}\left(\boldsymbol{x}^{l}\right)\boldsymbol{d}^{l}\leq-\eta\left\Vert \boldsymbol{d}^{l}\right\Vert ^{2}+\bar{f}_{i}^{l}\left(\bar{\boldsymbol{x}}^{l}\right)-\bar{f}_{i}^{l}\left(\boldsymbol{x}^{l}\right),\label{eq:ftdbound}
\end{equation}
where $\boldsymbol{d}^{l}=\bar{\boldsymbol{x}}^{l}-\boldsymbol{x}^{l}$,
and $\eta>0$ is some constant. From Assumption \ref{asm:convP},
the gradient of $f_{i}\left(\boldsymbol{x}\right)$ is Lipschitz continuous,
and thus there exists $L_{f}>0$ such that
\begin{align}
f_{i}\left(\boldsymbol{x}^{l+1}\right) & \leq f_{i}\left(\boldsymbol{x}^{l}\right)+\gamma^{l}\nabla^{T}f_{i}\left(\boldsymbol{x}^{l}\right)\boldsymbol{d}^{l}+L_{f}\left(\gamma^{l}\right)^{2}\left\Vert \boldsymbol{d}^{l}\right\Vert ^{2}\nonumber \\
 & =f\left(\boldsymbol{x}^{l}\right)+L_{f}\left(\gamma^{l}\right)^{2}\left\Vert \boldsymbol{d}^{l}\right\Vert ^{2}+f_{i}\left(\boldsymbol{x}^{l}\right)-f\left(\boldsymbol{x}^{l}\right)\nonumber \\
 & +\gamma^{l}\left(\nabla^{T}\bar{f}_{i}^{l}\left(\boldsymbol{x}^{l}\right)+\nabla^{T}f_{i}\left(\boldsymbol{x}^{l}\right)-\nabla^{T}\bar{f}_{i}^{l}\left(\boldsymbol{x}^{l}\right)\right)\boldsymbol{d}^{l}\nonumber \\
 & \overset{\textrm{a}}{\leq}f\left(\boldsymbol{x}^{l}\right)+f_{i}\left(\boldsymbol{x}^{l}\right)-f\left(\boldsymbol{x}^{l}\right)-\eta\gamma^{l}\left\Vert \boldsymbol{d}^{l}\right\Vert ^{2}\nonumber \\
 & +\gamma^{l}\left(\bar{f}_{i}^{l}\left(\bar{\boldsymbol{x}}^{l}\right)-\bar{f}_{i}^{l}\left(\boldsymbol{x}^{l}\right)\right)+o\left(\gamma^{l}\right)\nonumber \\
 & \leq f\left(\boldsymbol{x}^{l}\right)-\eta\gamma^{l}\left\Vert \boldsymbol{d}^{l}\right\Vert ^{2}+o\left(\gamma^{l}\right),\forall i=1,...,m,\label{eq:fxdecre}
\end{align}
where $o\left(\gamma^{l}\right)$ means that $\lim_{l\rightarrow\infty}o\left(\gamma^{l}\right)/\gamma^{l}=0$.
In (\ref{eq:fxdecre}-a), we used (\ref{eq:ftdbound}) and $\lim_{l\rightarrow\infty}\left\Vert \nabla^{T}f_{i}\left(\boldsymbol{x}^{l}\right)-\nabla^{T}\bar{f}_{i}^{l}\left(\boldsymbol{x}^{l}\right)\right\Vert =0$,
and the last inequality follows from $f_{i}\left(\boldsymbol{x}^{l}\right)\leq f\left(\boldsymbol{x}^{l}\right)$,
$\liminf_{l\rightarrow\infty}f\left(\boldsymbol{x}^{l}\right)-\bar{f}_{i}^{l}\left(\bar{\boldsymbol{x}}^{l}\right)\geq0$,
and $\lim_{l\rightarrow\infty}\left\Vert f_{i}\left(\boldsymbol{x}^{l}\right)-\bar{f}_{i}^{l}\left(\boldsymbol{x}^{l}\right)\right\Vert =0$.
Since (\ref{eq:fxdecre}) holds for all $i=1,...,m$, by choosing
a sufficiently large $l_{\epsilon}$, we have 
\begin{align}
f\left(\boldsymbol{x}^{l+1}\right)-f\left(\boldsymbol{x}^{l}\right) & \leq-\gamma^{l}\overline{\eta}\left\Vert \boldsymbol{d}^{l}\right\Vert ^{2}\nonumber \\
 & \leq-\gamma^{l}\overline{\eta}\epsilon^{'},\forall l\in\mathcal{T}_{\epsilon}^{'},\label{eq:gapf}
\end{align}
for some $\overline{\eta}>0$. Moreover, from Assumption \ref{asm:convP},
the directional derivative of $f\left(\boldsymbol{x}\right)$ is uniformly
bounded, and thus there exists a constant $C$ such that
\begin{equation}
\left|f\left(\boldsymbol{x}^{l+1}\right)-f\left(\boldsymbol{x}^{l}\right)\right|\leq C\left\Vert \boldsymbol{x}^{l+1}-\boldsymbol{x}^{l}\right\Vert \leq C^{'}\gamma^{l},\label{eq:gapf1}
\end{equation}
for some $C^{'}>0$. Finally, it follows from (\ref{eq:gapf}) and
(\ref{eq:gapf1}) that 
\begin{equation}
f\left(\boldsymbol{x}^{l}\right)\leq2\epsilon,\forall l\geq l_{\epsilon}.\label{eq:fxtbound}
\end{equation}
Since (\ref{eq:fxtbound}) is true for any $\epsilon>0$, it follows
that $\limsup_{l\rightarrow\infty}f\left(\boldsymbol{x}^{l}\right)\leq0$.

2. Then we prove that $\lim_{l\rightarrow\infty}\left\Vert \bar{\boldsymbol{x}}^{l}-\boldsymbol{x}^{l}\right\Vert =0,$
w.p.1. 

2.1: We first prove that $\liminf_{l\rightarrow\infty}\left\Vert \bar{\boldsymbol{x}}^{l}-\boldsymbol{x}^{l}\right\Vert =0$
w.p.1.

Note that the feasible problem in (\ref{eq:Pitert-1}) is strictly
convex, and thus the solution is uniquely given by $\bar{\boldsymbol{x}}^{l}$.
Therefore, when a feasible update is performed at iteration $l$,
we have $\bar{f}^{l}\left(\bar{\boldsymbol{x}}^{l}\right)\geq0$ and
\begin{align*}
\bar{\boldsymbol{x}}^{l}=\underset{\boldsymbol{x}}{\text{argmin}\:} & \bar{f}_{0}^{l}\left(\boldsymbol{x}\right)\\
s.t.\: & \bar{f}_{i}^{l}\left(\boldsymbol{x}\right)\leq\bar{f}^{l}\left(\bar{\boldsymbol{x}}^{l}\right),i=1,....,m.
\end{align*}
As a result, $\bar{\boldsymbol{x}}^{l}$ can be expressed in a unified
way as
\begin{align}
\bar{\boldsymbol{x}}^{l}=\underset{\boldsymbol{x}}{\text{argmin}}\: & \bar{f}_{0}^{l}\left(\boldsymbol{x}\right)\label{eq:Pixbart}\\
s.t.\: & \bar{f}_{i}^{l}\left(\boldsymbol{x}\right)\leq\nu^{l},i=1,....,m,\nonumber 
\end{align}
where $\nu^{l}=0$ when an objective update is performed and $\nu^{l}=\bar{f}^{l}\left(\bar{\boldsymbol{x}}^{l}\right)$
when a feasible update is performed. Since $\lim_{l\rightarrow\infty}\left|\bar{f}^{l}\left(\boldsymbol{x}^{l}\right)-f\left(\boldsymbol{x}^{l}\right)\right|=0$,
$\bar{f}^{l}\left(\bar{\boldsymbol{x}}^{l}\right)\leq\bar{f}^{l}\left(\boldsymbol{x}^{l}\right)$,
and we have proved that $\limsup_{l\rightarrow\infty}f\left(\boldsymbol{x}^{l}\right)\leq0$,
it follows that $\lim_{l\rightarrow\infty}\nu^{l}=0$. Let $\hat{\boldsymbol{x}}^{l}$
denote the projection of $\boldsymbol{x}^{l}$ on to the feasible
set of Problem (\ref{eq:Pixbart}). Then it follows from $\lim_{l\rightarrow\infty}\nu^{l}=0$,
$\limsup_{l\rightarrow\infty}\bar{f}^{l}\left(\boldsymbol{x}^{l}\right)=\limsup_{t\rightarrow\infty}f\left(\boldsymbol{x}^{l}\right)\leq0$,
and the strong convexity of $\bar{f}^{l}\left(\boldsymbol{x}^{l}\right)$
that 
\begin{equation}
\lim_{l\rightarrow\infty}\left\Vert \boldsymbol{x}^{l}-\hat{\boldsymbol{x}}^{l}\right\Vert =0.\label{eq:gapxhx}
\end{equation}
From the definition of $\bar{f}_{0}^{l}\left(\boldsymbol{x}\right)$
in (\ref{eq:upsurrgate}), $\bar{f}_{0}^{l}\left(\boldsymbol{x}\right)$
is uniformly strongly convex, and thus
\begin{align}
\nabla^{T}\bar{f}_{0}^{l}\left(\boldsymbol{x}^{l}\right)\boldsymbol{d}^{l} & \leq-\eta\left\Vert \boldsymbol{d}^{l}\right\Vert ^{2}+\bar{f}_{0}^{l}\left(\bar{\boldsymbol{x}}^{l}\right)-\bar{f}_{0}^{l}\left(\boldsymbol{x}^{l}\right)\nonumber \\
 & =-\eta\left\Vert \boldsymbol{d}^{l}\right\Vert ^{2}+\bar{f}_{0}^{l}\left(\bar{\boldsymbol{x}}^{l}\right)-\bar{f_{0}}^{l}\left(\hat{\boldsymbol{x}}^{l}\right)\nonumber \\
 & +\bar{f_{0}}^{l}\left(\hat{\boldsymbol{x}}^{l}\right)-\bar{f}_{0}^{l}\left(\boldsymbol{x}^{l}\right)\nonumber \\
 & \leq-\eta\left\Vert \boldsymbol{d}^{l}\right\Vert ^{2}+e\left(l\right),\label{eq:ftdbound-1}
\end{align}
for some $\eta>0$, where $\boldsymbol{d}^{l}=\bar{\boldsymbol{x}}^{l}-\boldsymbol{x}^{l}$,
$\lim_{l\rightarrow\infty}e\left(l\right)=0$, and the last equality
follows from (\ref{eq:gapxhx}). From Assumption \ref{asm:convP},
the gradient of $f_{0}\left(\boldsymbol{x}\right)$ is Lipschitz continuous,
and thus there exists $L_{0}>0$ such that
\begin{align*}
f_{0}\left(\boldsymbol{x}^{l+1}\right) & \leq f_{0}\left(\boldsymbol{x}^{l}\right)+\gamma^{l}\nabla^{T}f_{0}\left(\boldsymbol{x}^{l}\right)\boldsymbol{d}^{l}+L_{0}\left(\gamma^{l}\right)^{2}\left\Vert \boldsymbol{d}^{l}\right\Vert ^{2}\\
 & =f_{0}\left(\boldsymbol{x}^{l}\right)+L_{0}\left(\gamma^{l}\right)^{2}\left\Vert \boldsymbol{d}^{l}\right\Vert ^{2}\\
 & +\gamma^{l}\left(\nabla^{T}f_{0}\left(\boldsymbol{x}^{l}\right)-\nabla^{T}\bar{f_{0}}^{l}\left(\boldsymbol{x}^{l}\right)+\nabla^{T}\bar{f_{0}}^{l}\left(\boldsymbol{x}^{l}\right)\right)\boldsymbol{d}^{l}\\
 & \leq f_{0}\left(\boldsymbol{x}^{l}\right)-\gamma^{t}\eta\left\Vert \boldsymbol{d}^{l}\right\Vert ^{2}+o\left(\gamma^{l}\right),
\end{align*}
where in the last inequality, we used (\ref{eq:ftdbound-1}) and $\lim_{l\rightarrow\infty}\left\Vert \nabla^{T}f_{0}\left(\boldsymbol{x}^{l}\right)-\nabla^{T}\bar{f}_{0}^{l}\left(\boldsymbol{x}^{l}\right)\right\Vert =0$.
Let us show by contradiction that w.p.1. $\liminf_{l\rightarrow\infty}\left\Vert \bar{\boldsymbol{x}}^{l}-\boldsymbol{x}^{l}\right\Vert =0$.
Suppose $\liminf_{l\rightarrow\infty}\left\Vert \bar{\boldsymbol{x}}^{l}-\boldsymbol{x}^{l}\right\Vert \geq\chi>0$
with a positive probability. Then we can find a realization such that
$\left\Vert \boldsymbol{d}^{l}\right\Vert \geq\chi$ at the same time
for all $l$. We focus next on such a realization. By choosing a sufficiently
large $l_{0}$, there exists $\overline{\eta}>0$ such that
\begin{align}
f_{0}\left(\boldsymbol{x}^{l+1}\right)-f_{0}\left(\boldsymbol{x}^{l}\right) & \leq-\gamma^{l}\overline{\eta}\left\Vert \boldsymbol{d}^{l}\right\Vert ^{2},\forall l\geq l_{0}.\label{eq:gapf0xt}
\end{align}
It follows from (\ref{eq:gapf0xt}) that 
\[
f_{0}\left(\boldsymbol{x}^{l}\right)-f_{0}\left(\boldsymbol{x}^{l_{0}}\right)\leq-\overline{\eta}\chi^{2}\sum_{j=l_{0}}^{l}\left(\gamma^{j}\right)^{2},
\]
which, in view of $\sum_{j=l_{0}}^{\infty}\left(\gamma^{j}\right)^{2}=\infty$,
contradicts the boundedness of $\left\{ f_{0}\left(\boldsymbol{x}^{l}\right)\right\} $.
Therefore it must be $\liminf_{l\rightarrow\infty}\left\Vert \bar{\boldsymbol{x}}^{l}-\boldsymbol{x}^{l}\right\Vert =0$
w.p.1.

2.2: Then we prove that $\limsup_{l\rightarrow\infty}\left\Vert \bar{\boldsymbol{x}}^{l}-\boldsymbol{x}^{l}\right\Vert =0$
w.p.1.

We first prove a useful lemma.
\begin{lem}
\label{lem:gapxbar}There exists a constant $\hat{L}>0$ such that
\[
\left\Vert \bar{\boldsymbol{x}}^{l_{1}}-\bar{\boldsymbol{x}}^{l_{2}}\right\Vert \leq\hat{L}\left\Vert \boldsymbol{x}^{l_{1}}-\boldsymbol{x}^{l_{2}}\right\Vert +e\left(l_{1},l_{2}\right),
\]
where $\lim_{l_{1},l_{2}\rightarrow\infty}e\left(l_{1},l_{2}\right)=0$.
\end{lem}

\begin{IEEEproof}
From Lemma \ref{lem:fconv}, we have 
\begin{equation}
\left|\bar{f}_{i}^{l_{1}}\left(\boldsymbol{x}\right)-\bar{f}_{i}^{l_{2}}\left(\boldsymbol{x}\right)\right|\leq C\left\Vert \boldsymbol{x}^{l_{1}}-\boldsymbol{x}^{l_{2}}\right\Vert +e^{'}\left(l_{1},l_{2}\right),\label{eq:gapft1t2}
\end{equation}
for all $\boldsymbol{x}\in\mathcal{X}$ and $i=0,1,...,m$, where
$\lim_{l_{1},l_{2}\rightarrow\infty}e^{'}\left(l_{1},l_{2}\right)=0$.
Then it follows from (\ref{eq:gapft1t2}) and (\ref{eq:Pixbart}),
and the Lipschitz continuity and strong convexity of $\bar{f}_{i}^{l}\left(\boldsymbol{x}\right),\forall i$
that
\begin{equation}
\left\Vert \bar{\boldsymbol{x}}^{l_{1}}-\bar{\boldsymbol{x}}^{l_{2}}\right\Vert \leq C_{1}C\left\Vert \boldsymbol{x}^{l_{1}}-\boldsymbol{x}^{l_{2}}\right\Vert +C_{1}e^{'}\left(l_{1},l_{2}\right)+C_{2}\nu^{l}\label{eq:xt1t2}
\end{equation}
for some constant $C_{1},C_{2}>0$. Finally, Lemma \ref{lem:gapxbar}
follows from (\ref{eq:xt1t2}) immediately.
\end{IEEEproof}

Using Lemma \ref{lem:gapxbar} and following the same analysis as
that in \cite{Yang_TSP2016_SSCA}, Proof of Theorem 1, it can be shown
that \textbf{$\limsup_{l\rightarrow\infty}\left\Vert \bar{\boldsymbol{x}}^{l}-\boldsymbol{x}^{l}\right\Vert =0$
}w.p.1. This completes the proof.

\subsection{Proof of Theorem \ref{thm:Convergence-of-Algorithm1}\label{subsec:Proof-of-Theorem}}

According to Lemma \ref{lem:Convergence-surrogate}, Lemma \ref{lem:keylem},
and (\ref{eq:Pixbart}), $\boldsymbol{x}^{*}$ must be the optimal
solution of the following convex optimization problem almost surely:

\begin{align}
\underset{\boldsymbol{x}}{\text{min}}\: & \hat{f}_{0}\left(\boldsymbol{x}\right)\label{eq:Piterthead}\\
s.t.\: & \hat{f}_{i}\left(\boldsymbol{x}\right)\leq0,i=1,....,m.\nonumber 
\end{align}
Since the Slater condition is satisfied, the KKT condition of Problem
(\ref{eq:Piterthead}) implies that there exist $\lambda_{1},...,\lambda_{m}$
such that
\begin{align}
\nabla\hat{f}_{0}\left(\boldsymbol{x}\right)+\sum_{i}\lambda_{i}\nabla\hat{f}_{i}\left(\boldsymbol{x}^{*}\right) & =\boldsymbol{0},\nonumber \\
\hat{f}_{i}\left(\boldsymbol{x}^{*}\right) & \leq0,\:\forall i=1,...,m,\nonumber \\
\lambda_{i}\hat{f}_{i}\left(\boldsymbol{x}^{*}\right) & =0,\:\forall i=1,...,m.\label{KKTPhead}
\end{align}
It follows from Lemma \ref{lem:Convergence-surrogate} and (\ref{KKTPhead})
that $\boldsymbol{x}^{*}$ also satisfies the KKT condition of Problem
(\ref{eq:mainP}). This completes the proof.


\end{document}